 \documentclass[pmlr,twocolumn,10pt]{jmlr} % W&CP article

\usepackage{booktabs}
\usepackage{siunitx}

\theorembodyfont{\upshape}
\theoremheaderfont{\scshape}
\theorempostheader{:}
\theoremsep{\newline}

\jmlrvolume{LEAVE UNSET}
\jmlryear{2023}
\jmlrsubmitted{LEAVE UNSET}
\jmlrpublished{LEAVE UNSET}
\jmlrworkshop{Machine Learning for Health (ML4H) 2023} % W&CP title

\def\mathbi#1{\textbf{\em #1}}
\usepackage{cuted}
 \title[REMEDI]{REMEDI: REinforcement learning-driven adaptive MEtabolism modeling of primary sclerosing cholangitis DIsease progression}

\author{%
\Name{Chang Hu} \Email{changhu@illinois.edu}\\
\addr University of Illinois at Urbana-Champaign
\AND
\Name{Krishnakant V. Saboo} \Email{ksaboo2@illinois.edu}\\
\addr University of Illinois at Urbana-Champaign
\AND
\Name{Ahmad H. Ali} \Email{aliah@health.missouri.edu}\\
\addr University of Missouri School of Medicine
\AND
\Name{Brian D. Juran} \Email{Juran.Brian@mayo.edu}\\
\addr Mayo Clinic
\AND
\Name{Konstantinos N. Lazaridis} \Email{lazaridis.konstantinos@mayo.edu}\\
\addr Mayo Clinic
\AND
\Name{Ravishankar K. Iyer} \Email{rkiyer@illinois.edu}\\
\addr University of Illinois at Urbana-Champaign
}

\begin{document}

\maketitle

\begin{abstract}
Primary sclerosing cholangitis (PSC) is a rare disease wherein altered bile acid metabolism contributes to sustained liver injury. This paper introduces REMEDI, a framework that captures bile acid dynamics and the body's adaptive response during PSC progression that can assist in exploring treatments. REMEDI merges a differential equation (DE)-based mechanistic model that describes bile acid metabolism with reinforcement learning (RL) to emulate the body's adaptations to PSC continuously. An objective of adaptation is to maintain homeostasis by regulating enzymes involved in bile acid metabolism. These enzymes correspond to the parameters of the DEs. REMEDI leverages RL to approximate adaptations in PSC, treating homeostasis as a reward signal and the adjustment of the DE parameters as the corresponding actions. On real-world data, REMEDI generated bile acid dynamics and parameter adjustments consistent with published findings. Also, our results support discussions in the literature that early administration of drugs that suppress bile acid synthesis may be effective in PSC treatment.
\end{abstract}
\begin{keywords}
Reinforcement learning, Disease progression, Differential equation, Adaptation
\end{keywords}

\section{Introduction}
\label{ml4h_sec:introduction}
Primary sclerosing cholangitis (PSC) is a rare, complex liver disease in which altered bile acid metabolism contributes to liver injury \citep{ba_liver}. There are no effective medications, and liver transplantation is often necessary \citep{psc_therapy}. A critical hurdle in exploring therapeutics is the lack of a model capturing the relevant disease dynamics, the body's response to the disease, and the effects of treatments. We aim to develop a machine learning (ML) based PSC progression model with a focus on bile acid metabolism dynamics and its bidirectional interactions with the body over time. Such a model could facilitate treatment evaluations and accelerate drug discovery or repurposing. Examples of computational models guiding interventions already exist for prostate cancer \citep{zhang2017integrating} and HIV \citep{xiao2013modeling}.

There were three main challenges in developing the proposed progression model: (1) the absence of a bile acid metabolism model during PSC; (2) limited insight into the body's adaptive response to the disease; and (3) the lack of data from affected organs and a dearth of longitudinal data. (1) While prior studies have proposed differential equation (DE)-based bile acid metabolism models for healthy individuals \citep{sips_ba_model}, they do not capture bile duct obstruction, the pathophysiological hallmark of PSC \citep{psc_characteristics}, and its impact on bile acid metabolism. (2) Over the course of the disease, the body responds to changing bile acid levels by continually adapting and altering bile acid metabolism, which plays a central role in keeping PSC patients asymptomatic for many years \citep{psc_ascending_pathophysiology}. However, the specific adaptations during PSC progression are not well understood \citep{psc_adaptation_unclear}, making them difficult to model. (3) Despite the liver and the bile ducts being central to PSC, direct bile acid measurements in these organs are infeasible. We are limited to bile acid data in the blood. Moreover, these data are cross-sectional, i.e., taken only at a single time point, further complicating the modeling of longitudinal disease progression.

We introduce REMEDI to model PSC progression by extending existing bile acid metabolism DEs with PSC pathophysiology and incorporating a reinforcement learning (RL) agent to approximate the body's adaptations. REMEDI addresses the above challenges with the following key innovations:

(1) We developed a reduced-order bile acid metabolism model to capture dynamics pertinent to PSC, based on an existing DE model for healthy individuals. We extended the reduced model with clinical domain knowledge to capture bile duct obstruction. 

(2) We used RL to emulate the body's adaptive response to the disease. We assume the body is a smart agent that, through evolution, has learned to adapt itself to maintain homeostasis of critical metabolic events \citep{giordano2013homeostasis_evolution}. In PSC, the body regulates bile acid metabolism enzymes to maintain homeostasis. These enzymes are naturally represented as parameters in the reduced-order bile acid DE model. During disease progression, the RL agent constantly updates these parameters to maximize a reward function that promotes homeostasis and the generation of close-to-reality bile acid profiles. 

(3) Because PSC patients can have a prolonged, partially successful adaptation period \citep{psc_ascending_pathophysiology}, we assume the real-world cross-sectional data were collected during this ``stable" period, and we encourage the RL agent to generate stabilized trajectories that are close to the data.

The main assumption of REMEDI is that the goal of homeostasis drives biological actions, achieved by smart regulation of body enzymes \citep{savir2017homeostasis_enzyme, billman2020homeostasis_physiology_principle}. Thus, we treat adaptation as a sequential optimization problem with ``homeostasis" as the objective function and the sequential regulation of enzymes as the optimization arguments. These enzymes are represented as parameters in the DEs. RL offers a framework to solve this sequential optimization problem. In PSC, the bile acid DEs constitute the environment, ``homeostasis" is the reward function, and the modulation of DE parameters is the actions. Therefore, RL, in combination with the DEs, approximates the body's adaptation to disease and enables dynamic modeling of PSC progression.

\begin{figure*}[htbp]
\floatconts
  {fig:rl_overview}
  {\caption{Overview of the reinforcement learning formulation of bile acid metabolism adaptation in PSC.}}
  {\includegraphics[width=0.9\linewidth]{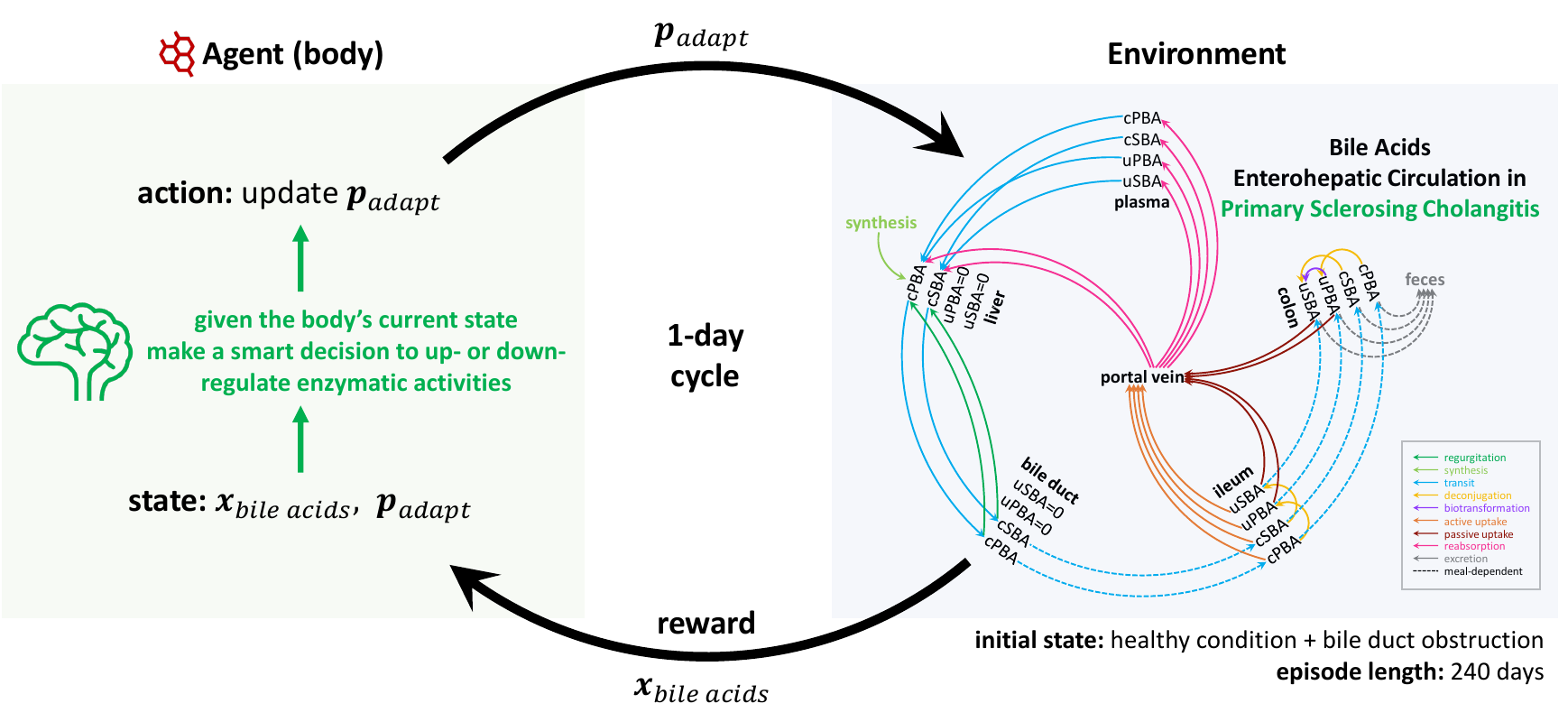}}
\end{figure*}

We validated REMEDI against findings from the literature and real-world clinical data. REMEDI produced biologically realistic results. (1) The reduced-order bile acid model captured the relevant dynamics of a more detailed model from the literature with drastically reduced computational cost. (2) Incorporation of bile duct obstruction mimicked PSC pathophysiology observed in clinical and animal studies \citep{psc_ascending_pathophysiology, bdl_intestine, bile_back_flow}. (3) On real-world PSC data, REMEDI generated bile acid dynamics and parameter adaptations consistent with the literature \citep{psc_ascending_pathophysiology}. (4) We evaluated \textit{in silico} the effects of two PSC drugs in clinical trials \citep{psc_therapy} and found REMEDI has the potential to explain the drugs' biologically observed behaviors \citep{PSC_potential_treatment, cabellerocamino2023asbt_inhibitor_improve_liver}.

Our contributions include: (1) developing the first mathematical model of bile acid metabolism in PSC, based on clinical domain knowledge, and (2) providing an \textit{in silico} testbed to evaluate the effects of bile acid modulating therapies. Our approach can be leveraged to determine optimal interventions for PSC in combination with comprehensive clinical data.

In principle, REMEDI's approach of using RL to estimate time-varying DE parameters can be extended to other diseases where DE-based models with time-varying parameters have been proposed, such as HIV \citep{hiv_ode}. Moreover, the innovative strategy of REMEDI that leverages RL to emulate adaptive behaviors holds promise for modeling a variety of homeostatic biological systems.

\section{REMEDI Framework}
\label{ml4h_sec:methods}
Our approach has three parts (\figureref{fig:rl_overview}): (1) a reduced-order bile acid metabolism model for healthy individuals (\sectionref{sec:method_healthy}), (2) a domain knowledge-based extension to depict the pathophysiology of PSC (\sectionref{sec:method_pathophysiology}), and (3) RL that captures the body’s adaptation to the pathophysiology (\sectionref{sec:method_adaptation}).

\subsection{Model of healthy bile acid metabolism}\label{sec:method_healthy}

Several species of bile acids circulate in multiple organs of the human body in a process called \textit{enterohepatic circulation} \citep{enterohepatoc_circulation}. In the liver (li), cholesterol is metabolized into the unconjugated primary bile acids cholic acid and chenodeoxycholic acid (uCA and uCDCA), which are transformed into conjugated forms (cCA and cCDCA) and secreted into the bile ducts and gallbladder (bd). These bile acids are then released into the intestines, where bacteria in the ileum (il) and colon (co) convert them into secondary bile acids (SBA). Active and passive uptake reabsorb intestinal bile acids back to the liver, with a small portion escaping into plasma (pl). Unreabsorbed bile acids are excreted with feces (fe).

We adopt \citet{sips_ba_model}'s approach and model bile acid metabolism with a series of DEs. Based on their relevance to PSC, we merge several bile acid species and do not distinguish among certain organ segments (see \appendixref{apd:ba_model}), resulting in a reduced-order model of bile acid $BA \in \{$cCA, cCDCA, cSBA, uCA, uCDCA, uSBA$\}$ in organ $OG \in \{$li, bd, il, co, pl, fe$\}$. For each BA in each OG, we (1) model its influxes/outfluxes from relevant biochemical and physical processes and (2) combine the fluxes into one DE to describe how the BA level in OG varies with time. See \appendixref{apd:ba_model} for all processes being modeled and \appendixref{apd:ba_model_equations} for all corresponding DEs.

\begin{figure}[htbp]
\floatconts
  {fig:ba_model_zoomed}
  {\caption{Fluxes affecting liver cCA level.}}
  {\includegraphics[width=0.75\linewidth]{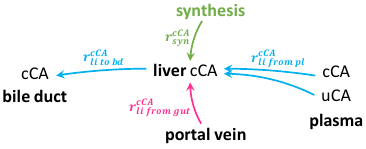}}
\end{figure}

As an example, \figureref{fig:ba_model_zoomed} shows the fluxes that affect the liver cCA level: (1) $r_{\text{syn}}^{\text{cCA}}$ is de novo synthesis in liver cells; (2) $r_{\text{li from gut}}^{\text{cCA}}$ is active and passive uptake from the gut; (3) $r_{\text{li from pl}}^{\text{cCA}}$ is influx from systemic blood circulation; and (4) $r_{\text{li to bd}}^{\text{cCA}}$ is outflux to the gallbladder and bile ducts. Hence, liver cCA level $x_{\text{li}}^{\text{cCA}}$ varies with time ($\mu$mol/min) according to:
\begin{equation}\label{eq:dx_li_cCA}
\frac{dx_{\text{li}}^{\text{cCA}}}{dt}=r_{\text{syn}}^{\text{cCA}}+r_{\text{li from gut}}^{\text{cCA}}+r_{\text{li from pl}}^{\text{cCA}}-r_{\text{li to bd}}^{\text{cCA}}
\end{equation}
\begin{equation}\label{eq:r_syn_cCA}
r_{\text{syn}}^{\text{cCA}} = p[\text{synthesis}] \cdot p[\text{syn\_frac\_CA}]
\end{equation}
\begin{equation}\label{eq:r_li_to_bd_cCA}
r_{\text{li to bd}}^{\text{cCA}} = p[\text{li\_to\_bd\_freq}] \cdot x_{\text{li}}^{\text{cCA}}
\end{equation}

We model all fluxes with zero- or first-order dynamics. For example, we model the cCA synthesis flux $r_{\text{syn}}^{cCA}$ as the product of two constant parameters, making $r_{\text{syn}}^{cCA}$ also a constant (zero-order). Here, $p[\text{synthesis}]$ denotes the total bile acid synthesis rate, and $p[\text{syn\_frac\_CA}]$ describes the fraction of cholic acid among the newly synthesized bile acids. The liver-to-bile duct transit rate of cCA, $r_{\text{li to bd}}^{\text{cCA}}$, exemplifies first-order dynamics. $x_{\text{li}}^{\text{cCA}}$ represents the current liver cCA level, and the constant parameter $p[\text{li\_to\_bd\_freq}]$ characterizes the first-order dynamics.

\subsection{Introducing PSC pathophysiology} \label{sec:method_pathophysiology}

\begin{figure*}[htbp]
\floatconts
  {fig:psc_pathophysiology}
  {\caption{Idealized modeling of PSC pathophysiology.}}
  {\includegraphics[width=0.9\linewidth]{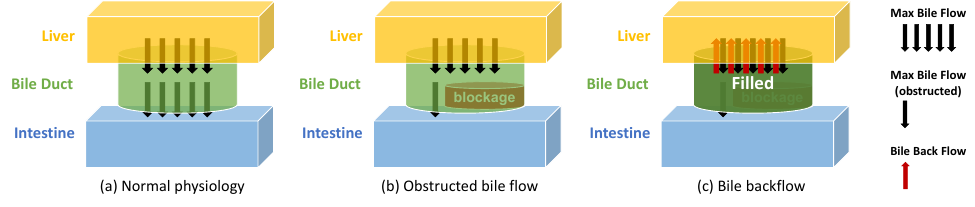}}
\end{figure*}

Drawing on clinical domain knowledge, we extend the reduced-order bile acid metabolism model with PSC pathophysiology (\figureref{fig:psc_pathophysiology}) by (1) implementing an obstruction of bile flow in the bile ducts and (2) introducing bile acid backflow to the liver following excessive bile acid buildup in the bile ducts.

\paragraph{Obstructed bile flow in the bile ducts:}
In PSC, chronic inflammation causes scarring and narrowing of the bile ducts \citep{psc_review}, impeding the normal bile flow into the small intestine (\figureref{fig:psc_pathophysiology}(b)). The extent of the obstruction determines the reduction of bile flow. We introduce a parameter $p[\text{bd\_max\_flow}]$ to denote the maximum amount of bile acids allowed to flow through, in proportion to the degree of obstruction. Consequently, if $r_{\text{bd to il}}$ calculated from its first-order dynamics exceeds $p[\text{bd\_max\_flow}]$, we cap it at $p[\text{bd\_max\_flow}]$:
\begin{equation}\label{eq:equation6}
r_{\text{bd to il}} = \min(r_{\text{bd to il}}, p[\text{bd\_max\_flow}])
\end{equation}

\paragraph{Bile acid backflow to the liver:}
The bile ducts (and gallbladder) have a limited storage capacity for bile acids. In PSC, bile duct obstruction can result in bile acid buildup exceeding the duct’s capacity, leading to regurgitation and backflow of excessive bile acids to the liver \citep{bile_back_flow}, as depicted in \figureref{fig:psc_pathophysiology}(c). To represent this, we introduce a parameter $p[\text{bd\_max\_ba}]$ denoting the bile acid holding capacity of the bile duct, and define $r_{\text{bd to li}}$ as the excessive bile acids backflowing from the bile duct to the liver:
\begin{equation}\label{eq:r_bd_to_li}
\begin{split}
    r_{\text{bd to li}} = \max(&x_{\text{bd}} + r_{\text{bd from li}} - r_{\text{bd to il}}\\
    &- p[\text{bd\_max\_ba}], 0)
\end{split}
\end{equation}

$x_{\text{bd}}$ is the current bile acid level in the bile duct, $r_{\text{bd from li}}$ is the influx from the liver, and $r_{\text{bd to il}}$ is the outflux to the ileum. See \appendixref{apd:ba_model_equations} for details.

\subsection{Model adaptation in PSC} \label{sec:method_adaptation}
We model adaptation as a series of enzymatic regulations to optimally maintain homeostasis, i.e., preserve physiological functions without deviating too much from the enzyme levels under healthy conditions. We model this adaptation process using RL (\figureref{fig:rl_overview}).

\paragraph{State:}
We assume the body self-regulates based on its current status, captured by the state vector $=\{\mathbi{x}_{bile\ acids}, \mathbi{p}_{adapt}\}$ in our context of bile acid metabolism. $\mathbi{x}_{bile\ acids}$ denotes bile acid levels across species and organs (30 variables, see \appendixref{apd:rl_setup}). $\mathbi{p}_{adapt}$ denotes parameter values corresponding to regulatable enzymes (five variables, see \textbf{Action}).

\paragraph{Action:}
We assume the body adapts through continual enzyme regulations, which translates to the modulation of the subset of DE parameters representing enzyme levels regulated by the body. Our DEs contain five such regulatable parameters, i.e., $\mathbi{p}_{adapt}=\{p[\text{synthesis}]\mathrel{,}p[\text{syn\_frac\_CA}]\mathrel{,}p[\text{hep\_ratio\_conj\_tri}]\mathrel{,}p[\text{hep\_ratio\_conj\_di}]\mathrel{,}p[\text{max\_asbt\_rate}]\}$. At every RL step, for each parameter in $\mathbi{p}_{adapt}$, any of three actions can be taken: up-regulation or down-regulation (with a prespecified fold change or absolute difference), or remaining unchanged. $p[\text{synthesis}]$ and $p[\text{syn\_frac\_CA}]$ were introduced in \sectionref{sec:method_healthy}. $p[\text{hep\_ratio\_conj\_tri}]$ denotes the fraction of reabsorbed bile acid extracted by the liver (in contrast to going into systemic blood) for cCA, and $p[\text{hep\_ratio\_conj\_di}]$ for cCDCA and cSBA. $p[\text{max\_asbt\_rate}]$ is the rate parameter in the first-order dynamics of active uptake.

\paragraph{Environment:}
The RL environment simulates bile acid dynamics following the introduction of PSC pathophysiology to healthy conditions. We use the reduced-order DEs extended with bile duct obstruction as the simulator. At every RL step, the RL agent modifies $\mathbi{p}_{adapt}$ in the DEs and updates the state vector. The environment takes a step forward via numerical integration of the DEs for a fixed duration. The resultant bile acid levels update $\mathbi{x}_{bile\ acids}$ in the state vector. A reward is computed in the environment step and sent to the RL agent to determine the next action. The simulation terminates when a prespecified time period is reached or when any state variable exceeds physiological ranges.

\paragraph{Reward:}
Our reward function comprises several terms to guide the RL agent towards meaningful adaptations that (1) sustain physiological functions (including minimizing liver toxicity, facilitating fat digestion, and maintaining cholesterol elimination), (2) resemble real-world patient data, and (3) conform with ranges and values reported in the literature.

\subparagraph{Minimizing toxicity.} One of the main goals of adaptation is to limit liver exposure ($LE$) to toxic bile acids \citep{PSC_potential_treatment}. We set a negative reward for excessive bile acid exposure in the liver to minimize liver toxicity. We calculate $LE$ as the cumulative liver bile acid level over one day. If current $LE$ exceeds $LE$ under healthy conditions, we set the negative reward to be the normalized excessive exposure.
\begin{equation}\label{eq:reward_toxicity}
-\max\left(\frac{\text{current $LE$}-\text{healthy $LE$}}{\text{maximum possible $LE$}}, 0\right)
\end{equation}

\subparagraph{Facilitating digestion.} Adaptation requires preserving digestive functions under disease conditions \citep{tappenden2014adaptation_digestion}. Because bile acids in the ileum are necessary for fat digestion, we promote ileum access ($IA$) to bile acids with a reward term defined as the ratio of current $IA$ to $IA$ under healthy conditions. $IA$ is calculated as the cumulative ileum bile acid level over one day. The reward is capped at 1, offering no additional benefit beyond healthy levels.
\begin{equation}\label{eq:reward_digestion}
\min\left(\frac{\text{current $IA$}}{\text{healthy $IA$}}, 1\right)
\end{equation}

\subparagraph{Maintaining cholesterol elimination.} Synthesizing bile acids from cholesterol is one of the main pathways for eliminating cholesterol from the body \citep{wang2018cholesterol_elimination_bile_acid_synthesis}. Insufficient elimination of cholesterol increases the risks of multiple diseases. We reward sufficient cholesterol elimination, represented by the ratio of the current bile acid synthesis ($BAS$) rate to the $BAS$ rate under healthy conditions. The reward is capped at 1, offering no additional benefit from excessive cholesterol elimination.
\begin{equation}
\min\left(\frac{\text{current $BAS$ rate}}{\text{healthy $BAS$ rate}}, 1\right)
\end{equation}

\subparagraph{Resembling real-world patient data.} The way to adapt and sustain physiological functions might not be unique. To obtain an RL agent that mirrors adaptation in humans, we set a reward term to promote RL solutions that resemble real-world plasma data from PSC patients. We select a representative patient from our cohort (see \sectionref{ml4h_sec:results}) and penalize the difference between the patient's data $BA_{data}$ and the respective RL states $BA_{RL}$. This difference is divided by the corresponding bile acid's standard deviation. The negative sum of the squared weighted difference is multiplied by a coefficient $\lambda\in[0,1]$ to match other reward terms’ range.
\begin{equation}\label{eq:reward_data}
-\lambda \min \left(\Sigma_{BA} \left(\frac{BA_{RL} - BA_{data}}{\text{std } BA_{data}}\right)^2, \text{CAP}\right)
\end{equation}
    
\subparagraph{Conforming with values from the literature:} We design additional reward terms to ensure the RL agent generates physiologically plausible bile acid levels while keeping the regulatable parameters close to their values under healthy conditions. See \appendixref{apd:rl_setup} for further details.

\subsection{Implementation of REMEDI}

\paragraph{Degree of pathophysiology:}
To determine the extent of bile duct obstruction, we ran a grid search for the parameter $p[\text{bd\_max\_flow}]$ across values of 1, 2, 3, 5, 10, 20, and 40 $\mu$mol/min, simulating scenarios from near-complete to partial obstruction. Corresponding RL agents were trained independently.

\paragraph{RL timeframes:}
At each step, the RL agent modified $\mathbi{p}_{adapt}$ to simulate the next 24 hours. Considering adaptation unfolds over days to weeks, this daily cycle offered sufficient opportunities for meaningful modulations. We restricted the simulation to a maximum of 240 days to adequately encapsulate the initial adaptation phase \citep{animal_model_time}.

\paragraph{Adaptation amplitudes:}
We chose relatively large parameter adaptation amplitudes to match the day-long RL steps. For parameters describing rates, i.e., $p[\text{synthesis}]$ and $p[\text{max\_asbt\_rate}]$,
we simulated up- or down-regulation with a 25\% higher or lower fold change; for parameters describing fractions, i.e.,  $p[\text{syn\_frac\_CA}]$, $p[\text{hep\_extract\_ratio\_conj\_tri}]$, and $p[\text{hep\_extract\_ratio\_conj\_di}]$), we applied a 10\% addition or subtraction. We also prespecified physiologically plausible ranges for each parameter, and up- or down-regulated values exceeding the ranges were clipped. The RL agent generated stochastic actions during training and evaluation.

\paragraph{Real-world patient data:} Our dataset includes plasma bile acid measurements of 222 PSC patients from our hospital partner. We selected five representative patients whose measurements minimized the sum of the distances to all patients. Patient-specific models were trained using their respective data, resulting in five models for the five patients. Since PSC is chronic, we assume patient data were collected during the stabilized adaptation phase. Hence, the sum of squared errors to encourage resemblance to patient data was only introduced after the initial four weeks.

\paragraph{RL training:}
We trained the RL agent for 4,000,000 environment steps using the model-free Proximal Policy Optimization (PPO) algorithm implemented in the Python package Stable Baselines3.

More details can be found in \appendixref{apd:rl_setup}.

\section{Results}
\label{ml4h_sec:results}
\subsection{Healthy bile acid dynamics}

We derived the DE parameters and initial bile acid values from the well-calibrated model by \citet{sips_ba_model}. We simulated our DEs over 60 days, reaching a steady state at which bile acid dynamics repeated every 24 hours. The Runge-Kutta 45 numerical solver was used to integrate the DEs in our experiments.

We validated our reduced-order DEs against the original DEs proposed by \citep{sips_ba_model} (see \appendixref{apd:ba_model} for details). The reduced-order DEs completed a 60-day simulation in 2.99s, over ten times faster than the 32.5s of the original DEs, while retaining similar bile acid trends (\appendixref{apd:ba_model} \figureref{fig:healthy_ba_24_hour}). The steady-state (day 60) bile acid levels of the reduced-order DEs were also consistent with real-world plasma data of 302 healthy individuals from our hospital partner (\appendixref{apd:ba_model} \figureref{fig:healthy_ba_univariate} and \figureref{fig:healthy_ba_bivariate}).

\begin{figure*}[!ht]
\floatconts
  {fig:psc_ba_comparison}
  {\caption{Simulated 60-day (panels (a)--(b)) and 240-day ((c)--(e)) bile acid dynamics after introduction of PSC pathophysiology ($p[\text{bd\_max\_flow}]$=3 $\mu$mol/min) to healthy conditions at day 0. (a) Without adaptation; (b)--(c) with adaptive DE parameters obtained from a trained RL agent; (d) with complete reduction of active uptake; (e) with 50\% reduction of active uptake. Results shown here were derived from the model trained with Patient 1's data. Only cCDCA values at 8 AM of each day are plotted. See \appendixref{apd:complete_ba} and \appendixref{apd:complete_ba_other_patients} for the complete dynamics from all patients.}}
  {\includegraphics[width=0.95\linewidth]{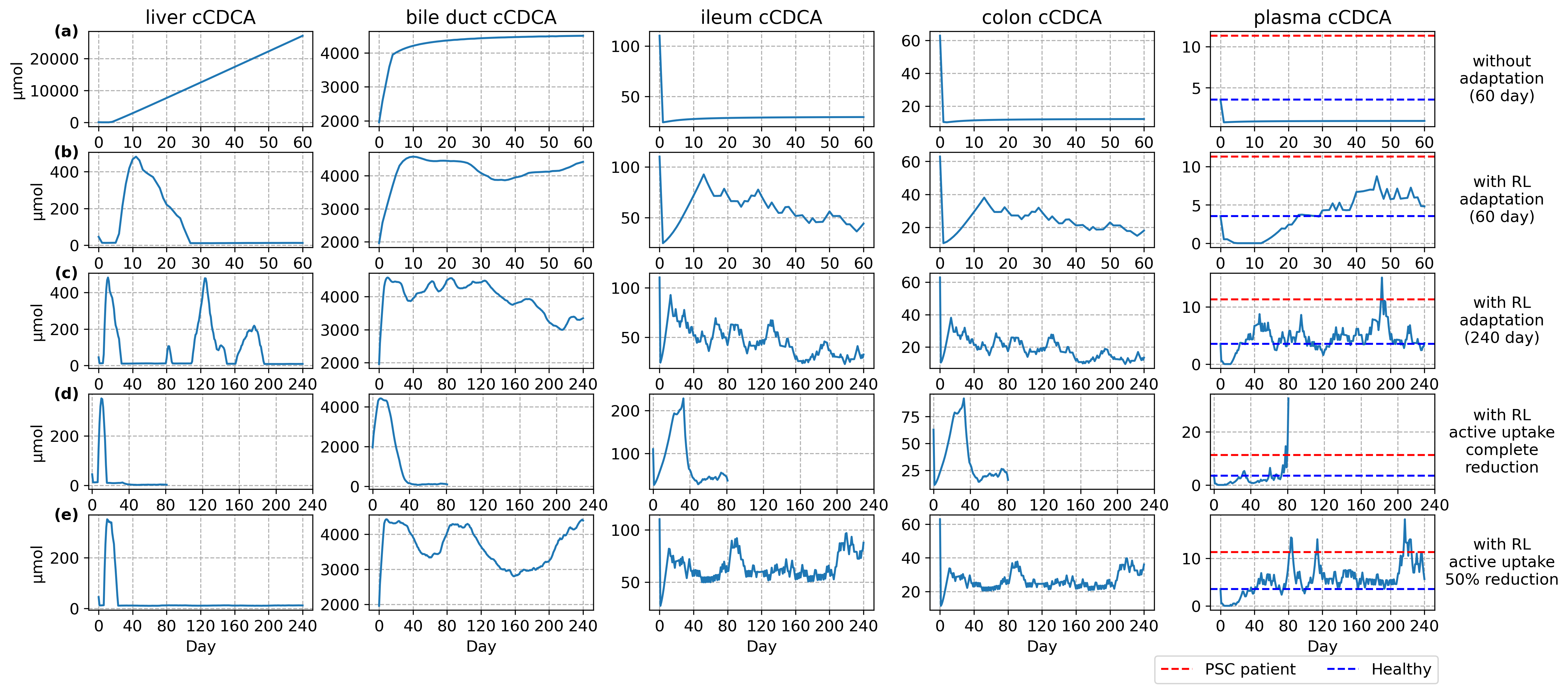}}
\end{figure*}

\subsection{PSC bile acid dynamics without RL}
For this analysis, we employed the reduced-order DEs extended with PSC pathophysiology, excluding RL adaptation. The parameter representing the degree of bile duct obstruction, $p[\text{bd\_max\_flow}]$, was set to 3 $\mu$mol/min (in contrast to 75 $\mu$mol/min when unobstructed). Other DE parameters and bile acid levels were initialized with their steady-state values under healthy conditions. We ran the simulation for 60 days, which, as shown by animal models, was enough time to establish adaptation \citep{animal_model_time}. 

\figureref{fig:psc_ba_comparison}(a) shows the 60-day cCDCA bile acid dynamics following the introduction of PSC pathophysiology (see \appendixref{apd:complete_ba} for dynamics of other bile acids). We observed a 230\% surge in bile duct bile acid levels, a direct result of impaired bile flow to the ileum and the subsequent accumulation in the bile ducts. A corresponding decrease in ileum bile acids was also observed. Around day five, bile duct bile acids reached a saturation point, causing excess bile acids to flow back into the liver. These observations align with the biologically expected changes \citep{bdl_intestine, bile_back_flow}.

However, once bile backflow starts, it continues at a near-constant rate throughout the remaining simulation, generating unrealistically high liver bile acid levels. The cCDCA level rose to 27,148 $\mu$mol by day 60, in stark contrast to the 45 $\mu$mol under healthy conditions. The simulation also indicated decreased conjugated bile acid levels in the plasma, conflicting with data showing elevated levels in PSC patients. These discrepancies arise from the flawed assumption that bile acid metabolism parameters remain unchanged, neglecting the dynamic adaptation occurring in PSC.

\subsection{PSC bile acid dynamics with REMEDI}

\begin{table*}[hbtp]
\floatconts
  {tab:pl_ba_difference}
  {\caption{Error between the model and PSC Patient 1's data ($\mu$mol, day 50 -- 60 averaged).}}
  {\begin{tabular}{lllllll}
  \toprule
  \bfseries Model/Plasma & \bfseries cCA & \bfseries cCDCA & \bfseries cSBA & \bfseries uCA & \bfseries uCDCA & \bfseries uSBA\\
  \midrule
    REMEDI without RL & -12.01 & -10.37 & -2.21 & -0.08 & -0.18 & -0.35\\
    % REMEDI (Patient 1) & -9.79 & -5.27 & 0.67 & 0.05 & 0.16 & 0.15\\
    REMEDI & -9.91 & -6.68 & 0.25 & 0.08 & 0.10 & 0.14\\
  \bottomrule
  \end{tabular}}
\end{table*}

We tested REMEDI upon introducing PSC pathophysiology to healthy conditions. We evaluated a range of $p[\text{bd\_max\_flow}]$ values and chose the case with $p[\text{bd\_max\_flow}]$=3 $\mu$mol/min for further analysis, as it yielded plasma bile acid levels closest to real data and therefore was more likely to reflect the real-world disease conditions (see \appendixref{apd:rl_setup}).

Upon introducing PSC pathophysiology in Patient 1's model, we observed an initial surge of bile acid levels in the bile ducts and a decrease in the downstream intestines (\figureref{fig:psc_ba_comparison}(b)), as observed in animal studies of PSC \citep{bdl_intestine}. Around day five, liver bile acids started to accumulate following saturation in the bile ducts. Importantly, REMEDI was able to adjust and stabilize the liver bile acid levels by week four, avoiding the unrealistic continuous rise seen without RL adaptation (\figureref{fig:psc_ba_comparison}(a)). Furthermore, REMEDI showed an increase in conjugated bile acids in the plasma and a decrease in unconjugated forms (\appendixref{apd:complete_ba} \figureref{fig:psc_ba_w_rl}), better aligning with the real-world PSC patient data than without RL adaptation (\figureref{fig:psc_ba_comparison}(c)). Similar results were seen for models trained on Patient 2--5 (\appendixref{apd:complete_ba_other_patients}).

To quantitatively assess the improvement from RL, we compared Patient 1's measurements with bile acid predictions from models trained on data from four other patients. The average error with RL was smaller than the error without RL (\tableref{tab:pl_ba_difference}). We adopted this approach for validation as we only had one cross-sectional measurement for each patient.

Overall, REMEDI with RL-based adaptation captured key disease dynamics and generated a more faithful representation of real-world data.

\subsection{Trajectories of adaptive parameters}
Analyzing parameter adjustments by REMEDI and linking them to the underlying enzymes can shed light on possible adaptive mechanisms driving bile acid metabolism in PSC. Notably, bile acid synthesis went through a sharp decline following the bile duct obstruction and remained low throughout the simulation period (\figureref{fig:psc_ode_param_w_rl}), in line with the known down-regulation of the bile acid synthesis enzyme CYP7A1 in cholestasis (a condition in which the bile flow from the liver stops or slows) \citep{psc_ascending_pathophysiology}. 

The model also predicted fluctuations in the CA:CDCA ratio among newly synthesized bile acids ($p[\text{syn\_frac\_CA}]$). CA:CDCA ratio depends on the activities of the classical and alternative bile acid synthesis pathways, which are regulated by enzymes such as CYP8B1 \citep{li2012ca_cdca_ratio_cyp8b1}. How the activities of these enzymes change in PSC is unclear, but our real-world PSC patient data also showed an altered plasma CA:CDCA ratio, warranting further study of potential enzymatic shifts. 

We observed reduced liver extraction of reabsorbed intestinal bile acids, with more bile acids entering systemic blood ($p[\text{hep\_extract\_ratio\_conj\_tri}]$ and $p[\text{hep\_extract\_ratio\_conj\_di}]$), potentially explaining the elevated plasma bile acid levels commonly seen in PSC patients. This trend aligns with cholestasis-related down-regulation of NTCP, a key enzyme in liver bile acid extraction \citep{donner2007cholestasis_ntcp_downregulation}.

Finally, there was a temporary drop in the ileum bile acid active uptake efficiency in the initial adaptation phase ($p[\text{max\_asbt\_rate}]$), corroborating the down-regulation of the bile acid uptake mediating enzyme ASBT in animal studies \citep{hruz2006cholestasis_asbt_downregulation}.

Trends in Patient 2--5 were similar (\appendixref{apd:complete_ba_other_patients}).

\begin{figure}[htbp]
\floatconts
  {fig:psc_ode_param_w_rl}
  {\caption{DE parameter adaptations in REMEDI trained on Patient 1's data.}}
  {\includegraphics[width=1\linewidth]{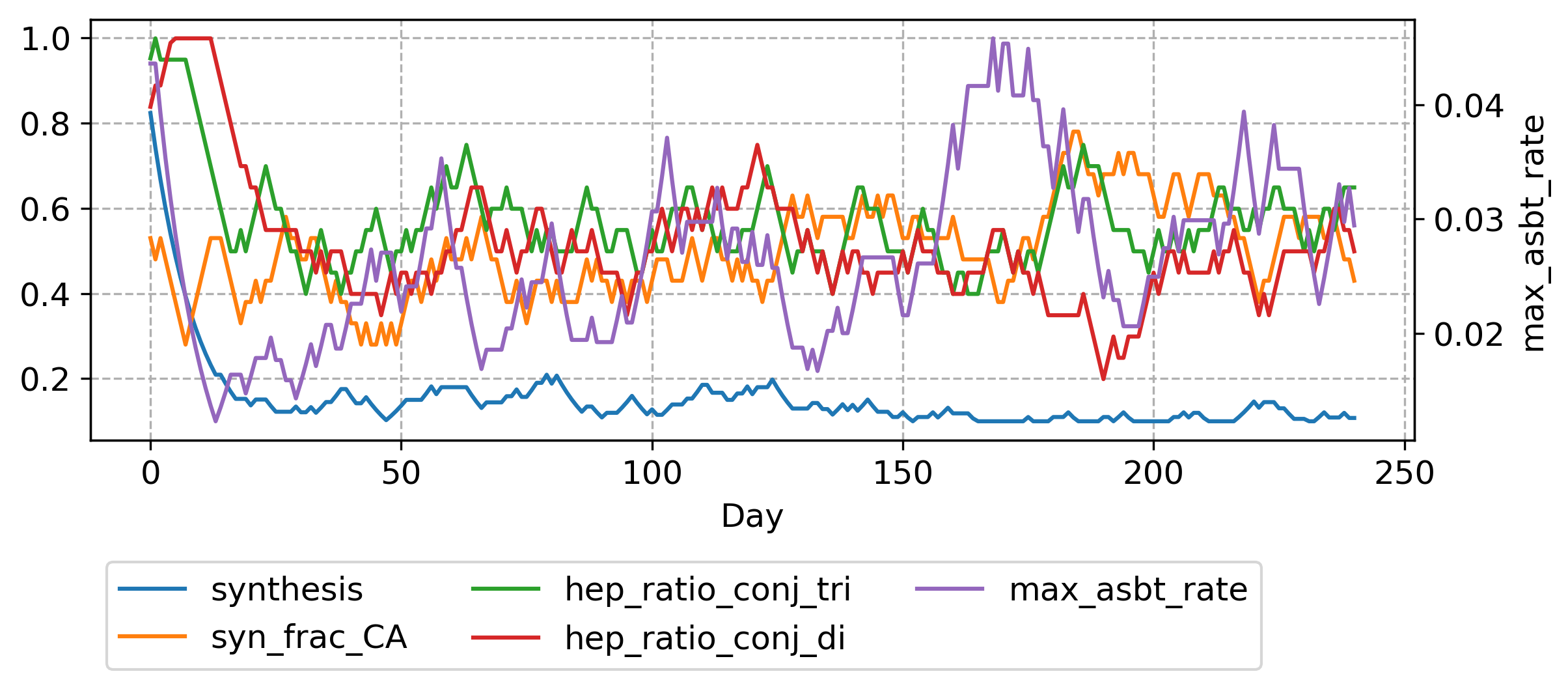}}
\end{figure}

\subsection{\textit{In silico} evaluation of bile acid therapies}
Using the trained REMEDI model, we assessed two types of experimental therapies targeting bile acid metabolism \textit{in silico} \citep{PSC_potential_treatment}: (1) suppression of bile acid synthesis, which we studied through analyzing bile acid synthesis rate ($p[\text{synthesis}]$) adjustments during adaptation, and (2) reduction of bile acid active uptake, which we simulated by decreasing efficiency of intestinal active uptake ($p[\text{max\_asbt\_rate}]$). Our analysis suggested partial reduction of bile acid active uptake might protect the liver, highlighting the value of REMEDI in evaluating therapies \textit{in silico}.

\paragraph{Suppressing synthesis:}
Drugs suppressing bile acid synthesis enzyme CYP7A1 are in clinical trials for PSC and other cholestatic conditions \citep{chiang2020cyp7a1_therapy}. Interestingly, REMEDI suggested the liver may naturally inhibit bile acid synthesis ($p[\text{synthesis}]$) as an adaptive response (\figureref{fig:psc_ba_comparison}(c)), implying the therapeutic window for such drugs may be limited to the early stages of the disease, before endogenous adaptations establish.

\paragraph{Reducing active uptake:}
ASBT, a central enzyme in ileum bile acid active uptake, has been the target of multiple drugs \citep{psc_therapy}. We simulated two strategies targeting this mechanism by adjusting $p[\text{max\_asbt\_rate}]$: (1) complete reduction, down-regulating $p[\text{max\_asbt\_rate}]$ until it reaches its physiological lower bound, and (2) 50\% reduction, down-regulating $p[\text{max\_asbt\_rate}]$ until it reaches or falls below half of the lower bound.

\subparagraph{Complete reduction:}
REMEDI predicted that full inhibition of active uptake would result in physiologically implausible spikes in plasma and intestine bile acids (\figureref{fig:psc_ba_comparison}(d)), which led to premature termination of the simulation, implying complete reduction is likely an unrealistic strategy.

\subparagraph{50\% reduction:} Partial reduction limited the occurrence of liver bile acid spikes (\figureref{fig:psc_ba_comparison}(e)), potentially protecting the liver from excessive bile acid accumulation, in line with animal studies showing a liver-protection effect from an ASBT inhibitor \citep{cabellerocamino2023asbt_inhibitor_improve_liver}. In contrast to complete reduction, 50\% reduction yielded plasma and intestine bile acid levels within realistic ranges, making partial reduction a more viable strategy.

\section{Related Works}
\label{ml4h_sec:relatedwork}
Several studies have proposed DE-based mechanistic models of bile acid metabolism \citep{hofmann1983ba_model,molino1986ba_model,sips_ba_model, baier2019ba_model, voronova2020ba_model}, albeit for healthy or non-PSC conditions. Moreover, the adaptive responses of the body in pathological conditions were usually not considered. A notable exception is \citet{voronova2020ba_model}, which explicitly modeled the FXR-FGF19 bile acid self-regulation pathway \citep{eloranta2008fxr_in_ba}. Our approach is unique for using RL to simultaneously consider multiple regulation pathways without explicitly modeling their mechanisms. A previous study combined DEs with RL to model Alzheimer's disease progression \citep{saboo2021reinforcement}. They used RL to estimate DE variable values that maximize cognition and minimize energetic cost, whereas we estimate DE parameters representing enzyme levels that promote homeostasis.

\section{Limitations}
\label{ml4h_sec:limitations}
First, we made several assumptions due to the current limited understanding of PSC, including focusing on bile acid metabolism, modeling bile duct obstruction as a sudden blockage, and in setting adaptation goals. Refining these assumptions as new insights of PSC emerge will lead to a more realistic model. Second, our bile acid trajectory prediction was only compared to a single time point because we only had access to cross-sectional data. Future collection of longitudinal data and data from multiple organs will be crucial for further validating REMEDI.

\section{Conclusion}
\label{ml4h_sec:conclusion}
We developed REMEDI, a novel model of PSC progression by combining bile acid metabolism DEs with an RL agent that captures the body's adaptation. REMEDI captures key bile acid trends in disease progression consistent with the literature and predicted therapy responses \textit{in silico}.

% \acks{}

\bibliography{jmlr-sample}

\appendix

\clearpage
\section{Reduced-Order Bile Acid Metabolism Model Under Healthy Conditions}\label{apd:ba_model}
\subsection{Bile Acids Enterohepatic Circulation}
\begin{figure}[htbp]
\floatconts
  {fig:ba_model_overview}
  {\caption{Overview of the biochemical and physical processes modeled in the bile acids enterohepatic circulation. We use primary bile acids (PBA) to represent both cholic acid (CA) and chenodeoxycholic acid (CDCA) because they share the same processes.}}
  {\includegraphics[width=1\linewidth]{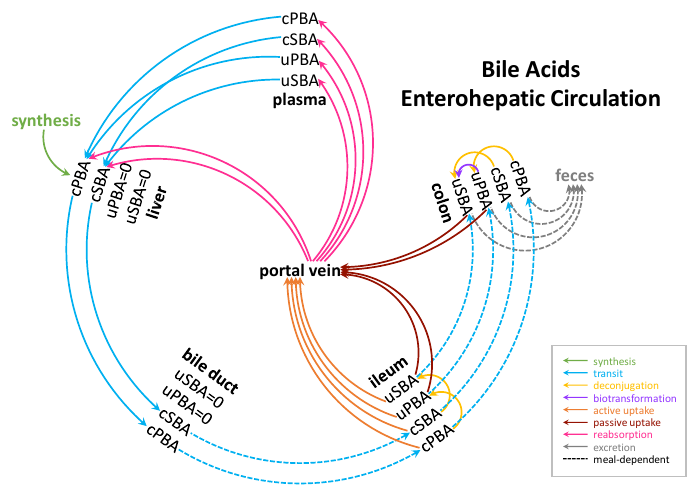}}
\end{figure}

Several species of bile acids circulate in multiple organs of the human body in a process called enterohepatic circulation. In the liver (li), cholesterol is metabolized into primary bile acids unconjugated cholic acid (uCA), and unconjugated chenodeoxycholic acid (uCDCA), which are then conjugated with either glycine or taurine into conjugated cholic acid (cCA) and conjugated chenodeoxycholic acid (cCDCA). Conjugated primary bile acids are then secreted into the bile duct and gallbladder (bd), which store and secrete the bile acids into the intestine, mostly after meals. Bacteria in the ileum (il) and the colon (co) deconjugate some of these bile acids and colon bacteria dehydroxylate primary bile acids to form secondary bile acids (SBA). Through active and passive uptake, bile acids are reabsorbed into the portal vein and then back to the liver. Active uptake happens in the ileum, while passive uptake happens in both the ileum and the colon, but only for unconjugated bile acids. Liver cells extract the reabsorbed bile acids efficiently, with a small portion of various bile acid species escaping into the systemic circulation (plasma, pl). Bile acids that are not reabsorbed in the intestines are excreted with feces (fe).

\subsection{Model Reduction from \citet{sips_ba_model}}
\citet{sips_ba_model} presents a detailed bile acid model that is calibrated with a comprehensive dataset compiled from 99 studies. For each bile acid (BA) in each organ (OG), \citet{sips_ba_model} (1) models its influxes/outfluxes from relevant biochemical and physical processes and (2) joins the fluxes into one differential equation to describe how the level of BA in OG varies with time. We adopt \citet{sips_ba_model}'s approach and model bile acid circulation with a series of ordinary differential equations (ODEs). Based on relevance to PSC, we reduce \citet{sips_ba_model}'s model by merging several bile acid species and not distinguishing certain organ segments. Specifically, \citet{sips_ba_model} divided the intestines into 15 consecutive segments for a detailed description of postprandial dynamics, which we deem irrelevant for PSC and combine the segments the ileum and the colon; \citet{sips_ba_model} modeled 8 different cSBA species separately, while we combine them together because SBA as a whole is hypothesized to play a role in PSC (\citet{secondary_ba}). Our reduction results in the modeling of bile acid $BA \in \{$cCA, cCDCA, cSBA, uCA, uCDCA, uSBA$\}$ in organ $OG \in \{$li, bd, il, co, pl, fe$\}$. \figureref{fig:ba_model_overview} summarizes the biochemical and physical processes in the enterohepatic circulation that constitute the ODEs.

\subsection{Meal-related Dynamics}
In addition to the zero and first-order dynamics discussed in the main text, our ODEs also contain time-varying parameters that capture the nuances of meal-related dynamics. When we eat, the gallbladder contracts and releases bile acids, and the bowels also move faster. As a result, bile acids move through these compartments at an increased pace immediately after eating, gradually slowing down over time. To model these meal-dependent dynamics, we assume the parameters governing the first-order dynamics of the transit rates along these compartments are functions of the time passed since the last meal, parametrized by normalized Rayleigh functions. Specifically, meal-dependent dynamics affect transit from the gallbladder and bile duct to the ileum, from the ileum to the colon, and from the colon to feces. As an example, 

\begin{equation}\label{eq:r_co_to_fe_cCA}
r_{\text{co to fe}}^{\text{cCA}} = p[\text{co\_to\_fe\_freq}] \cdot p[\text{co\_transit\_coef}] \cdot x_{\text{co}}^{\text{cCA}}
\end{equation}

\begin{equation}\label{eq:p_co_transit_coef}
\begin{split}
    &p[\text{co\_transit\_coef}] = \\
    &1+p[\text{peak\_co}]\cdot\left(\frac{t_{\text{meal}}\sqrt{e}}{p[\text{loc\_co}]}\exp{\left(-\frac{t_{\text{meal}}}{2\cdot p[\text{loc\_co}]^2}\right)}\right)   
\end{split}
\end{equation}

$t_{\text{meal}}$ represents the time that has passed since the subject's most recent meal. As a result of the normalized Rayleigh function, $p[\text{co\_transit\_coef}]$ starts to rise from 1 immediately after a meal, reaching its postprandial peak of $1+p[\text{peak\_co}]$ at time $p[\text{loc\_co}]$. Subsequently, it gradually returns to 1.

\subsection{Estimates of Model Parameters and Fasting Bile Acid Levels}
We utilize the model fitting and simulation results in \citet{sips_ba_model} to derive the parameters and fasting bile acid levels for a healthy individual for our model. For example, since $p[\text{synthesis}]$ has the same definition in \citet{sips_ba_model} as in our model, we directly take its value from \citet{sips_ba_model}. However, there is no equivalent variable in \citet{sips_ba_model} for $x_{\text{co}}^{\text{cCA}}$, which represents the level of conjugated cholic acid in the colon. Instead, \citet{sips_ba_model} separates conjugated cholic acid into taurine-conjugated cholic acid and glycine-conjugated cholic acid and models their levels in 5 consecutive parts of the colon. Therefore, we derive our fasting $x_{\text{co}}^{\text{cCA}}$ by summing up taurine-conjugated cholic acid and glycine-conjugated cholic acid in the 5 parts of the colon using the fasting time simulation results of \citet{sips_ba_model}. Similarly, we estimate fasting $r_{\text{co to fe}}^{\text{cCA}}$ by summing up the fluxes of taurine-conjugated cholic acid and glycine-conjugated cholic acid that transit from the last part of the colon to feces. Because $p[\text{co\_transit\_coef}]$ is assumed to be close to 1 at fasting, we derive $p[\text{co\_to\_fe\_freq}]$ by dividing fasting $r_{\text{co to fe}}$ by fasting $x_{\text{co}}$ and taking the average across bile acid species. Other parameters and bile acid levels are similarly estimated.

\subsection{Specification of Bile Acid Dynamics Simulation under Healthy Conditions}
After obtaining the estimations for all parameters and fasting bile acid levels, we run the system of ODEs using the Runge-Kutta 45 (RK45) numerical solver. To mimic the meal-dependent characteristics of bile acid metabolism, we introduce three simulated meals at 0, 6, and 12 hours every 24 hours, corresponding to breakfast, lunch, and dinner (8 AM, 2 PM, and 8 PM). We simulate the ODE system for a duration of 60 days, surpassing the 55-day duration in \citet{sips_ba_model}. We verify that a steady state has been reached by day 60, where the bile acids dynamics approximately repeat themselves every 24 hours. Specifically, the largest relative change in a single bile acid is below 0.01\% between two consecutive days and remains below 0.1\% after a year (360 days). We then take the steady-state fasting bile acid levels as the fasting bile acid levels representative of healthy individuals.

\begin{figure*}[!ht]
\floatconts
  {fig:healthy_ba_24_hour}
  {\caption{Simulated 24-hour bile acid dynamics with meals at 8 AM, 2 PM, and 8 PM. Orange curve: dynamics from our reduced-order ODEs; blue curve: dynamics from \citet{sips_ba_model}.}}
  {\includegraphics[width=1\linewidth]{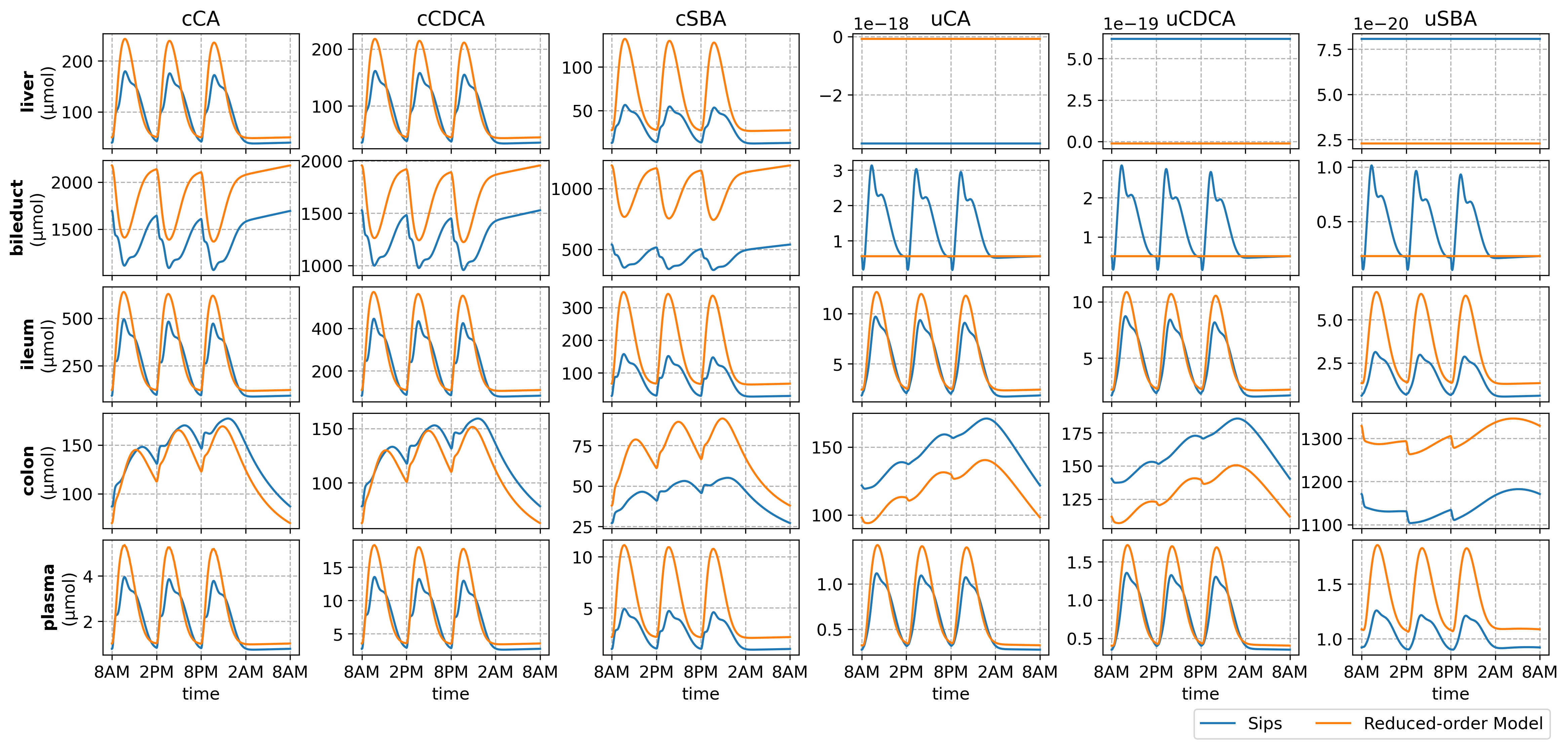}}
\end{figure*}

\subsection{24-hour Bile Acid Dynamics}
\figureref{fig:healthy_ba_24_hour} displays the 24-hour bile acid dynamics with three simulated meals at 8 AM, 2 PM, and 8 PM. The simulated results show that different bile acid species in different compartments all exhibit three-peak patterns that are associated with the three simulated meals. However, as expected, the peaking patterns and peaking amplitudes differ across bile acid species and compartments, validating the importance of modeling them separately. In general, conjugated bile acids in the same compartment follow similar trends, as do unconjugated bile acids, except for uSBA in the colon. Consistent with experimental observations, our results show that the gallbladder and bile duct contains the highest amount of conjugated bile acids among all compartments; uSBA is primarily concentrated in the colon; and plasma bile acid levels are generally low for all bile acid species due to the liver’s effective hepatic clearance under healthy conditions. Our simulation results also show good correspondence with the well-calibrated \citet{sips_ba_model} while reducing the computation time by more than 10X.

\subsection{Validation with A Real-world Dataset}
We validate the accuracy of our model by comparing the model simulation results with a real-world dataset from our partnering hospital. The dataset collected fasting plasma bile acid levels from 302 healthy individuals. Measurements below the detectable threshold were imputed with half of the smallest non-zero value. We take the model plasma bile acid levels at 8 AM, 12 hours after the previous simulated meal, as the corresponding fasting levels for comparison. Both the model simulated results and the real-world data were log 10 transformed. \figureref{fig:healthy_ba_univariate} shows that the simulated fasting bile acid levels are located within high likelihood regions of real-world data for all bile acids species, indicating that our model is capable of generating realistic bile acid levels. To further evaluate the representativeness of our simulation results beyond univariate comparison, we estimated the joint distribution of pairs of bile acids from the healthy control data and plotted the simulated data on the same plot in \figureref{fig:healthy_ba_bivariate}. We see the simulated fasting profile is still located within high-likelihood regions when pairs of bile acids are considered simultaneously. These validation results provide confidence in the model’s ability to simulate bile acid dynamics under healthy conditions and encourage its usage as a base model for studying study bile acid-related diseases.

\begin{figure}[htbp]
\floatconts
  {fig:healthy_ba_univariate}
  {\caption{log10 fasting plasma bile acid level (red vertical line) from the reduced-order ODEs compared to real-world data histograms.}}
  {\includegraphics[width=1\linewidth]{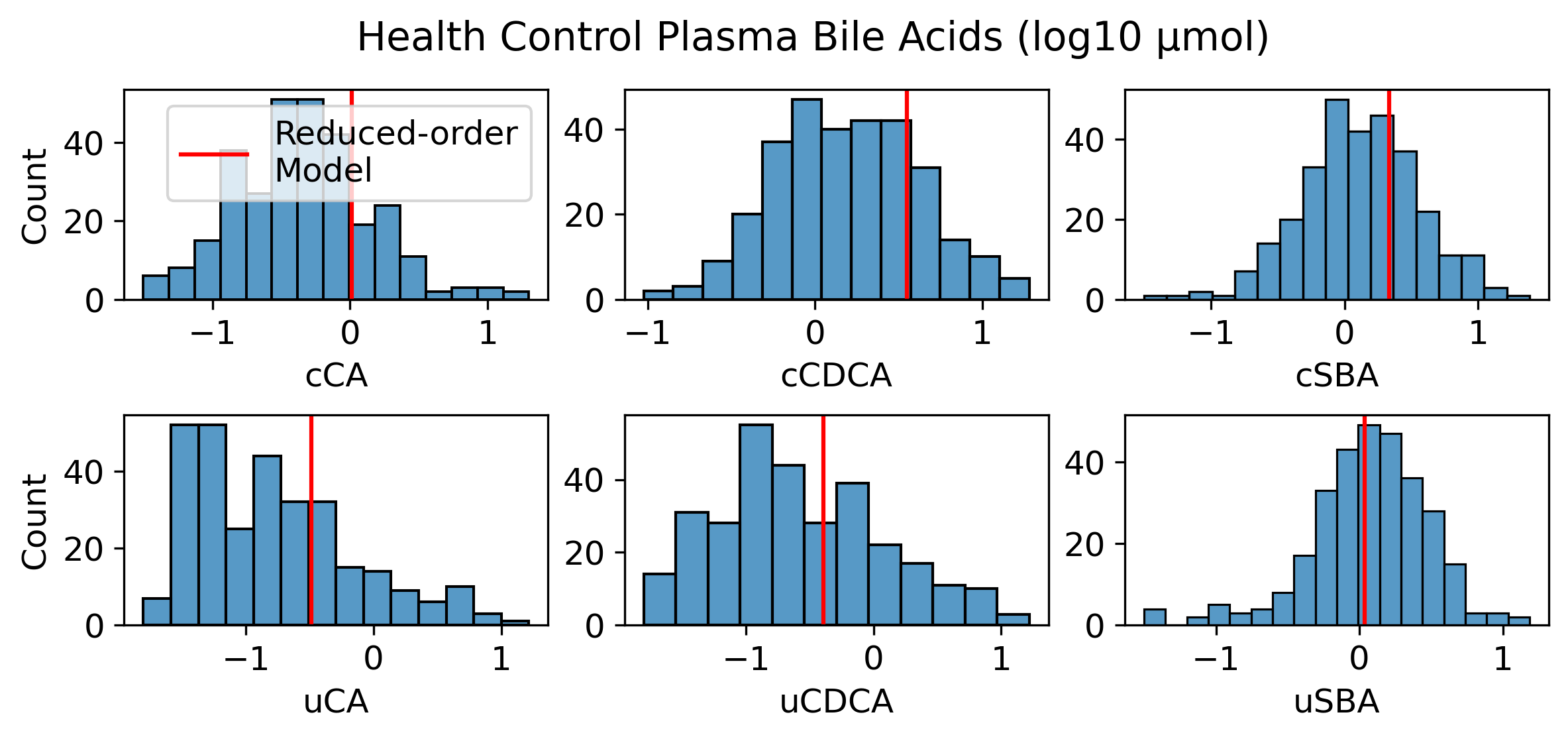}}
\end{figure}

\begin{figure}[htbp]
\floatconts
  {fig:healthy_ba_bivariate}
  {\caption{log10 fasting plasma level of bile acid pairs (red dot) from the reduced-order ODEs compared to real-world data bile acid pairs joint distribution.}}
  {\includegraphics[width=1\linewidth]{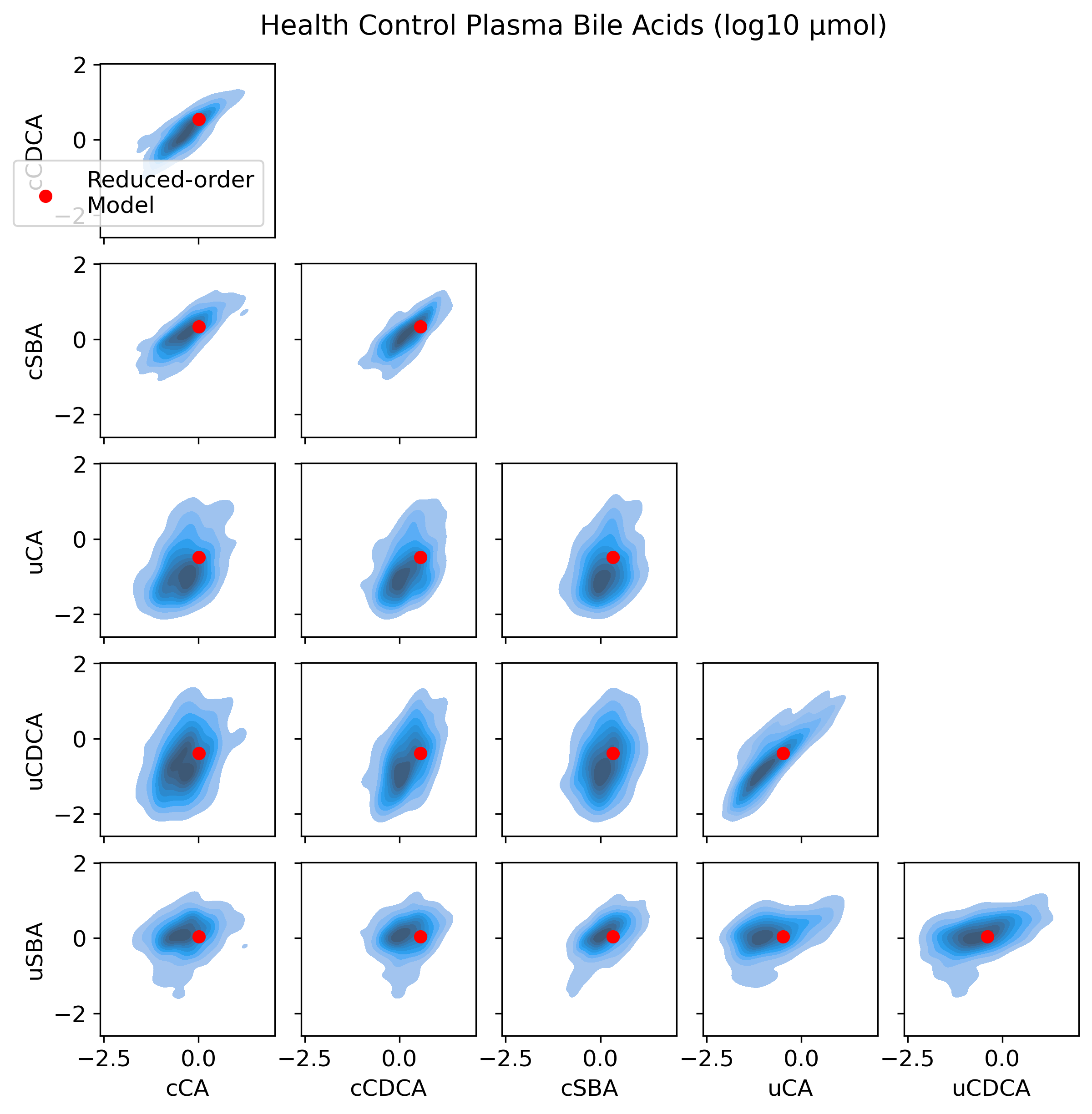}}
\end{figure}

\clearpage
\begingroup
    \allowdisplaybreaks
    \begin{strip}
\section{Reduced-Order Bile Acid Metabolism ODEs for PSC (Mathematical Descriptions)}\label{apd:ba_model_equations}

\subsection{Abbreviations}
\begin{itemize}
    \item cCA: conjugated cholic acid
    \item cCDCA: conjugated chenodeoxycholic acid
    \item cSBA: conjugated secondary bile acid
    \item uCA: unconjugated cholic acid
    \item uCDCA: unconjugated chenodeoxycholic acid
    \item uSBA: unconjugated secondary bile acid
    \item li: liver
    \item bd: bild duct
    \item il: ileum
    \item co: colon
    \item pl: plasma
    \item fe: feces
    \item syn: synthesis
    \item deconj: deconjugation
    \item biotr: primary to secondary bile acids bio-transformation
    \item au: active uptake
    \item pu: passive uptake
\end{itemize}

\subsection{Definition of independent parameters}
\begin{itemize}
    \item $p[\text{synthesis}]$: $\mu$mol/min. Total bile acid (BA) synthesis (cCA, cCDCA) rate in li.
    \item $p[\text{syn\_frac\_CA}]$ $\in[0,1]$, unitless. Fraction of CA among the newly synthesized BA.
    \item $p[\text{li\_to\_bd\_freq}]$: 1/min. Parameter for BA secretion from li to bd.
    \item $p[\text{bd\_to\_il\_freq}]$: 1/min. Parameter for BA transit from bd to il.
    \item $p[\text{il\_to\_co\_freq}]$: 1/min. Parameter for BA transit from il to co.
    \item $p[\text{co\_to\_fe\_freq}]$: 1/min. Parameter for BA excretion from co to fe.
    \item $p[\text{pl\_volume}]$: L. Total volume of pl in the body.
    \item $p[\text{pl\_flow\_thru\_li}]$: L/min. Rate of pl flowing through li.
    \item $p[\text{hep\_ratio\_conj\_tri}]$ $\in[0,1]$, unitless. For conjugated trihydroxylated BA (cCA) extracted from the gut via passive uptake and active uptake, the proportion that goes to li; for conjugated trihydroxylated BA (cCA) flowing through li from pl, the proportion that goes to li.
    \item $p[\text{hep\_ratio\_conj\_di}]$ $\in[0,1]$, unitless. Same as $p[\text{hep\_ratio\_conj\_tri}]$, but for conjugated dihydroxylated BA (primary BA CDCA, secondary BA ursodeoxycholic acid and deoxycholic acid). For simplicity, we assume the other secondary BA, mono-hydroxylated BA lithocholic acid, also belongs to this category. Hence, $p[\text{hep\_ratio\_conj\_di}]$ applies to cCDCA and cSBA.
    \item $p[\text{hep\_ratio\_unconj\_to\_conj}]$ $\in[0,1]$, unitless. Unconjugated BAs have lower hepatic extraction ratios than conjugated BAs. This parameter quantifies the fraction $\frac{p[\text{hep\_ratio\_unconj\_tri/di}]}{p[\text{hep\_ratio\_conj\_tri/di}]}$.
    \item $p[\text{gut\_deconj\_freq\_co}]$: 1/min. Parameter for BA deconjugation in co (by bacteria).
    \item $p[\text{gut\_deconj\_freq\_il\_to\_co}]$ $\in[0,1]$, unitless. Deconjugation happens more in co than in il. This parameter quantifies the fraction $\frac{p[\text{gut\_deconj\_freq\_il}]}{p[\text{gut\_deconj\_freq\_co}]}$.
    \item $p[\text{gut\_biotr\_freq\_CA}]$: 1/min. Parameter for uCA to uSBA biotransformation in co (by bacteria).
    \item $p[\text{gut\_biotr\_freq\_CDCA\_to\_CA}]$ $\in[0,1]$, unitless. Biotransformation to secondary BA happens (slightly) faster for CA than for CDCA. This parameter quantifies the fraction $\frac{p[\text{gut\_biotr\_freq\_CDCA}]}{p[\text{gut\_biotr\_freq\_CA}]}$.
    \item $p[\text{max\_asbt\_rate}]$: 1/min. Parameter for BA active uptake in il (via enzyme ASBT).
    \item $p[\text{gut\_pu\_freq\_co}]$: 1/min. Parameter for uBA passive uptake in co.
    \item $p[\text{gut\_pu\_freq\_il\_to\_co}]$ $\in[0,1]$, unitless. Passive uptake is more efficient in co than in il because of the larger diameter of co. This parameter quantifies the fraction $\frac{p[\text{gut\_pu\_freq\_il}]}{p[\text{gut\_pu\_freq\_co}]}$.
    \item $p[\text{meal\_reflex\_loc\_bd}]$: min. Parameter characterizing the normalized Rayleigh function that describes the meal-dependent BA transit dynamics from bd to il. BA transit efficiency from bd to il peaks at $p[\text{meal\_reflex\_loc\_bd}]$ min after the most recent meal, while baseline efficiency happens immediately after a meal or after an infinitely extended period.
    \item $p[\text{meal\_reflex\_loc\_il}]$: min. Same as $p[\text{meal\_reflex\_loc\_bd}]$, but for il to co transit.
    \item $p[\text{meal\_reflex\_loc\_co}]$: min. Same as $p[\text{meal\_reflex\_loc\_bd}]$, but for co to fe transit.
    \item $p[\text{meal\_reflex\_peak\_bd}]$: unitless. Parameter characterizing the normalized Rayleigh function that describes the meal-dependent BA transit dynamics from bd to il. At its peak, BA transit efficiency from bd to il is elevated by a factor of $1+p[\text{meal\_reflex\_peak\_bd}]$ compared to the baseline. 

    \item $p[\text{meal\_reflex\_peak\_il}]$: min. Same as $p[\text{meal\_reflex\_peak\_bd}]$, but for il to co transit.
    \item $p[\text{meal\_reflex\_peak\_co}]$: min. Same as $p[\text{meal\_reflex\_peak\_bd}]$, but for co to fe transit.
    \item $p[\text{co\_sulfate\_freq}]$: 1/min. Parameter for sulfation of secondary BA (LCA). For simplicity, we account for sulfation by modeling its end outcome: the additional excretion of SBA into fe, which is again modeled with first-order dynamics, governed by $p[\text{co\_sulfate\_freq}]$.
    \item $p[\text{bd\_max\_ba}]$: $\mu$mol. BA holding capacity of bd. Excessive BAs will backflow to li proportional to their levels in bd.
    \item $p[\text{bd\_max\_flow}]$: $\mu$mol. Maximum amount of BA allowed to pass through bd to il, in proportion to the degree of bd obstruction.
\end{itemize}

\subsection{ODE joining the bile acid fluxes}
How the level of each bile acid $BA \in \{$cCA, cCDCA, cSBA, uCA, uCDCA, uSBA$\}$ in each organ $OG \in \{$li, bd, il, co, pl, fe$\}$ varies with time is described by differential equations summing the corresponding influxes and subtracting the outfluxes. The fluxes take unit $\mu$mol/min unless specified otherwise.
\subsubsection{Liver}
\begin{alignat*}{2}
    &\frac{dx_{\text{li}}^{\text{cCA}}}{dt} &&= r_{\text{syn}}^{\text{cCA}}+r_{\text{li from gut}}^{\text{cCA}}+r_{\text{li from pl}}^{\text{cCA}}-r_{\text{li to bd}}^{\text{cCA}}+r_{\text{li from bd}}^{\text{cCA}}\\
    &\frac{dx_{\text{li}}^{\text{cCDCA}}}{dt} &&= r_{\text{syn}}^{\text{cCDCA}}+r_{\text{li from gut}}^{\text{cCDCA}}+r_{\text{li from pl}}^{\text{cCDCA}}-r_{\text{li to bd}}^{\text{cCDCA}}+r_{\text{li from bd}}^{\text{cCDCA}}\\
    &\frac{dx_{\text{li}}^{\text{cSBA}}}{dt} &&= r_{\text{li from gut}}^{\text{cSBA}}+r_{\text{li from pl}}^{\text{cSBA}}-r_{\text{li to bd}}^{\text{cSBA}}+r_{\text{li from bd}}^{\text{cSBA}}\\
    &\frac{dx_{\text{li}}^{\text{uCA}}}{dt} &&= 0\\
    &\frac{dx_{\text{li}}^{\text{uCDCA}}}{dt} &&= 0\\
    &\frac{dx_{\text{li}}^{\text{uSBA}}}{dt} &&= 0
\end{alignat*}
\subsubsection{Bile duct}
\begin{alignat*}{2}
    &\frac{dx_{\text{bd}}^{\text{cCA}}}{dt} &&= r_{\text{bd from li}}^{\text{cCA}}-r_{\text{bd to il}}^{\text{cCA}}-r_{\text{bd to li}}^{\text{cCA}}\\
    &\frac{dx_{\text{bd}}^{\text{cCDCA}}}{dt} &&= r_{\text{bd from li}}^{\text{cCDCA}}-r_{\text{bd to il}}^{\text{cCDCA}}-r_{\text{bd to li}}^{\text{cCDCA}}\\
    &\frac{dx_{\text{bd}}^{\text{cSBA}}}{dt} &&= r_{\text{bd from li}}^{\text{cSBA}}-r_{\text{bd to il}}^{\text{cSBA}}-r_{\text{bd to li}}^{\text{cSBA}}\\
    &\frac{dx_{\text{bd}}^{\text{uCA}}}{dt} &&= 0\\
    &\frac{dx_{\text{bd}}^{\text{uCDCA}}}{dt} &&= 0\\
    &\frac{dx_{\text{bd}}^{\text{uSBA}}}{dt} &&= 0
\end{alignat*}
\subsubsection{Ileum}
\begin{alignat*}{2}
    &\frac{dx_{\text{il}}^{\text{cCA}}}{dt} &&= r_{\text{il from bd}}^{\text{cCA}}-r_{\text{il deconj}}^{\text{cCA}}-r_{\text{il au}}^{\text{cCA}}-r_{\text{il to co}}^{\text{cCA}}\\
    &\frac{dx_{\text{il}}^{\text{cCDCA}}}{dt} &&= r_{\text{il from bd}}^{\text{cCDCA}}-r_{\text{il deconj}}^{\text{cCDCA}}-r_{\text{il au}}^{\text{cCDCA}}-r_{\text{il to co}}^{\text{cCDCA}}\\
    &\frac{dx_{\text{il}}^{\text{cSBA}}}{dt} &&= r_{\text{il from bd}}^{\text{cSBA}}-r_{\text{il deconj}}^{\text{cSBA}}-r_{\text{il au}}^{\text{cSBA}}-r_{\text{il to co}}^{\text{cSBA}}\\
    &\frac{dx_{\text{il}}^{\text{uCA}}}{dt} &&= r_{\text{il deconj}}^{\text{uCA}}-r_{\text{il au}}^{\text{uCA}}-r_{\text{il pu}}^{\text{uCA}}-r_{\text{il to co}}^{\text{uCA}}\\
    &\frac{dx_{\text{il}}^{\text{uCDCA}}}{dt} &&= r_{\text{il deconj}}^{\text{uCDCA}}-r_{\text{il au}}^{\text{uCDCA}}-r_{\text{il pu}}^{\text{uCDCA}}-r_{\text{il to co}}^{\text{uCDCA}}\\
    &\frac{dx_{\text{il}}^{\text{uSBA}}}{dt} &&= r_{\text{il deconj}}^{\text{uSBA}}-r_{\text{il au}}^{\text{uSBA}}-r_{\text{il pu}}^{\text{uSBA}}-r_{\text{il to co}}^{\text{uSBA}}
\end{alignat*}
\subsubsection{Colon}
\begin{alignat*}{2}
    &\frac{dx_{\text{co}}^{\text{cCA}}}{dt} &&= r_{\text{co from il}}^{\text{cCA}}-r_{\text{co deconj}}^{\text{cCA}}-r_{\text{co to fe}}^{\text{cCA}}\\
    &\frac{dx_{\text{co}}^{\text{cCDCA}}}{dt} &&= r_{\text{co from il}}^{\text{cCDCA}}-r_{\text{co deconj}}^{\text{cCDCA}}-r_{\text{co to fe}}^{\text{cCDCA}}\\
    &\frac{dx_{\text{co}}^{\text{cSBA}}}{dt} &&= r_{\text{co from il}}^{\text{cSBA}}-r_{\text{co deconj}}^{\text{cSBA}}-r_{\text{co to fe}}^{\text{cSBA}}-r_{\text{co sulfate}}^{\text{cSBA}}\\
    &\frac{dx_{\text{co}}^{\text{uCA}}}{dt} &&= r_{\text{co from il}}^{\text{uCA}}+r_{\text{co deconj}}^{\text{uCA}}-r_{\text{co biotr}}^{\text{uCA}}-r_{\text{co pu}}^{\text{uCA}}-r_{\text{co to fe}}^{\text{uCA}}\\
    &\frac{dx_{\text{co}}^{\text{uCDCA}}}{dt} &&= r_{\text{co from il}}^{\text{uCDCA}}+r_{\text{co deconj}}^{\text{uCDCA}}-r_{\text{co biotr}}^{\text{uCDCA}}-r_{\text{co pu}}^{\text{uCDCA}}-r_{\text{co to fe}}^{\text{uCDCA}}\\
    &\frac{dx_{\text{co}}^{\text{uSBA}}}{dt} &&= r_{\text{co from il}}^{\text{uSBA}}+r_{\text{co deconj}}^{\text{uSBA}}+r_{\text{co biotr}}^{\text{uSBA}}-r_{\text{co pu}}^{\text{uSBA}}-r_{\text{co to fe}}^{\text{uSBA}}-r_{\text{co sulfate}}^{\text{uSBA}}
\end{alignat*}
\subsubsection{Plasma}
\begin{alignat*}{2}
    &\frac{dx_{\text{pl}}^{\text{cCA}}}{dt} &&= r_{\text{pl from gut}}^{\text{cCA}}-r_{\text{pl to li}}^{\text{cCA}}\\
    &\frac{dx_{\text{pl}}^{\text{cCDCA}}}{dt} &&= r_{\text{pl from gut}}^{\text{cCDCA}}-r_{\text{pl to li}}^{\text{cCDCA}}\\
    &\frac{dx_{\text{pl}}^{\text{cSBA}}}{dt} &&= r_{\text{pl from gut}}^{\text{cSBA}}-r_{\text{pl to li}}^{\text{cSBA}}\\
    &\frac{dx_{\text{pl}}^{\text{uCA}}}{dt} &&= r_{\text{pl from gut}}^{\text{uCA}}-r_{\text{pl to li}}^{\text{uCA}}\\
    &\frac{dx_{\text{pl}}^{\text{uCDCA}}}{dt} &&= r_{\text{pl from gut}}^{\text{uCDCA}}-r_{\text{pl to li}}^{\text{uCDCA}}\\
    &\frac{dx_{\text{pl}}^{\text{uSBA}}}{dt} &&= r_{\text{pl from gut}}^{\text{uSBA}}-r_{\text{pl to li}}^{\text{uSBA}}\\
\end{alignat*}
\subsubsection{Feces}
\begin{alignat*}{2}
    &\frac{dx_{\text{fe}}^{\text{cCA}}}{dt} &&= r_{\text{fe from co}}^{\text{cCA}}\\
    &\frac{dx_{\text{fe}}^{\text{cCDCA}}}{dt} &&= r_{\text{fe from co}}^{\text{cCDCA}}\\
    &\frac{dx_{\text{fe}}^{\text{cSBA}}}{dt} &&= r_{\text{fe from co}}^{\text{cSBA}}+r_{\text{fe sulfate}}^{\text{cSBA}}\\
    &\frac{dx_{\text{fe}}^{\text{uCA}}}{dt} &&= r_{\text{fe from co}}^{\text{uCA}}\\
    &\frac{dx_{\text{fe}}^{\text{uCDCA}}}{dt} &&= r_{\text{fe from co}}^{\text{uCDCA}}\\
    &\frac{dx_{\text{fe}}^{\text{uSBA}}}{dt} &&= r_{\text{fe from co}}^{\text{uSBA}}+r_{\text{fe sulfate}}^{\text{uSBA}}
\end{alignat*}

\subsection{Helper functions and parameters}
We define the following helper functions and parameters as functions of the estimated parameters:
\subsubsection{Helpers for bile duct obstruction and bile acid backflow}
\begin{alignat*}{2}
    &x_{\text{bd}} &&= x_{\text{bd}}^{\text{cCA}} + x_{\text{bd}}^{\text{cCDCA}} + x_{\text{bd}}^{\text{cSBA}}\\
    &r_{\text{bd from li}} && = r_{\text{bd from li}}^{\text{cCA}} + r_{\text{bd from li}}^{\text{cCDCA}} + r_{\text{bd from li}}^{\text{cSBA}}\\
    &r_{\text{bd to il}} && = r_{\text{bd to il}}^{\text{cCA}} + r_{\text{bd to il}}^{\text{cCDCA}} + r_{\text{bd to il}}^{\text{cSBA}})\\    
    &p[\text{bdl\_discount}] &&= \min\left(\frac{p[\text{bd\_max\_flow}]}{p[\text{bd\_to\_il\_freq}] \cdot p[\text{bd\_transit\_coef}] \cdot x_{\text{bd}}}, 1\right)\\
    &p[\text{bd\_backflow\_coef}] &&= \max\left(\frac{x_{\text{bd}} + r_{\text{bd from li}} - r_{\text{bd to il}} - p[\text{bd\_max\_ba}]}{x_{\text{bd}}}, 0\right)
\end{alignat*}

\subsubsection{Helpers for parameters defined by ratios}
\begin{alignat*}{2}
    &p[\text{hep\_ratio\_unconj\_tri}] &&= p[\text{hep\_ratio\_conj\_tri}] \cdot p[\text{hep\_ratio\_unconj\_to\_conj}]\\
    &p[\text{hep\_ratio\_unconj\_di}] &&= p[\text{hep\_ratio\_conj\_di}] \cdot p[\text{hep\_ratio\_unconj\_to\_conj}]\\
    &p[\text{gut\_deconj\_freq\_il}] &&= p[\text{gut\_deconj\_freq\_co}] \cdot p[\text{gut\_deconj\_freq\_il\_to\_co}]\\
    &p[\text{gut\_biotr\_freq\_CDCA}] &&= p[\text{gut\_biotr\_freq\_CA}] \cdot p[\text{gut\_biotr\_freq\_CDCA\_to\_CA}]\\
    &p[\text{gut\_pu\_freq\_il}] &&= p[\text{gut\_pu\_freq\_co}] \cdot p[\text{gut\_pu\_freq\_il\_to\_co}]\\
\end{alignat*}

\subsubsection{Helpers for meal-related dynamics}
\begin{alignat*}{2}
    &\text{GI\_reflex}(loc,peak,t) &&= 1 + peak \cdot \left(\frac{t\sqrt{e}}{loc}\exp{\left(-\frac{t^2}{2\cdot loc^2}\right)}\right)\\
    &\text{GI\_reflex}(loc,peak,0) &&= 1\\
    &\text{GI\_reflex}(loc,peak,loc) &&= 1 + peak\\
    &\text{GI\_reflex}(loc,peak,\infty) &&= 1\\
    &p[\text{bd\_transit\_coef}] &&= \text{GI\_reflex}(p[\text{meal\_reflex\_loc\_bd}],p[\text{meal\_reflex\_peak\_bd}],t\_since\_meal)\\
    &p[\text{il\_transit\_coef}] &&= \text{GI\_reflex}(p[\text{meal\_reflex\_loc\_il}],p[\text{meal\_reflex\_peak\_il}],t\_since\_meal)\\
    &p[\text{co\_transit\_coef}] &&= \text{GI\_reflex}(p[\text{meal\_reflex\_loc\_co}],p[\text{meal\_reflex\_peak\_co}],t\_since\_meal)
\end{alignat*}

\subsection{Zero- and first-order dynamics defining the fluxes}
\subsubsection{Synthesis fluxes}
\begin{alignat*}{2}
    &r_{\text{syn}}^{\text{cCA}} &&= p[\text{synthesis}] \cdot p[\text{syn\_frac\_CA}]\\
    &r_{\text{syn}}^{\text{cCDCA}} &&= p[\text{synthesis}] \cdot (1-p[\text{syn\_frac\_CA}])
\end{alignat*}

\subsubsection{Transit fluxes}
Transit fluxes are further categorized based on the organs involved in the transit process:
\paragraph{Liver vs. Bile Duct}
\begin{alignat*}{2}
    &r_{\text{li to bd}}^{\text{cCA}} &&= p[\text{li\_to\_bd\_freq}] \cdot x_{\text{li}}^{\text{cCA}}\\
    &r_{\text{li to bd}}^{\text{cCDCA}} &&= p[\text{li\_to\_bd\_freq}] \cdot x_{\text{li}}^{\text{cCDCA}}\\
    &r_{\text{li to bd}}^{\text{cSBA}} &&= p[\text{li\_to\_bd\_freq}] \cdot x_{\text{li}}^{\text{cSBA}}\\
    &r_{\text{bd from li}}^{\text{cCA}} &&= r_{\text{li to bd}}^{\text{cCA}}\\
    &r_{\text{bd from li}}^{\text{cCDCA}} &&= r_{\text{li to bd}}^{\text{cCDCA}}\\
    &r_{\text{bd from li}}^{\text{cSBA}} &&= r_{\text{li to bd}}^{\text{cSBA}}\\
    &r_{\text{bd to li}}^{\text{cCA}} &&= p[\text{bd\_backflow\_coef}] \cdot x_{\text{bd}}^{\text{cCA}}\\
    &r_{\text{bd to li}}^{\text{cCDCA}} &&= p[\text{bd\_backflow\_coef}] \cdot x_{\text{bd}}^{\text{cCDCA}}\\
    &r_{\text{bd to li}}^{\text{cSBA}} &&= p[\text{bd\_backflow\_coef}] \cdot x_{\text{bd}}^{\text{cSBA}}\\
    &r_{\text{li from bd}}^{\text{cCA}} &&= r_{\text{bd to li}}^{\text{cCA}}\\
    &r_{\text{li from bd}}^{\text{cCDCA}} &&= r_{\text{bd to li}}^{\text{cCDCA}}\\
    &r_{\text{li from bd}}^{\text{cSBA}} &&= r_{\text{bd to li}}^{\text{cSBA}}
\end{alignat*}
\paragraph{Bile Duct vs. Ileum}
\begin{alignat*}{2}
    &r_{\text{bd to il}}^{\text{cCA}} &&= p[\text{bd\_discount}] \cdot p[\text{bd\_to\_il\_freq}] \cdot p[\text{bd\_transit\_coef}] \cdot x_{\text{bd}}^{\text{cCA}}\\
    &r_{\text{bd to il}}^{\text{cCDCA}} &&= p[\text{bd\_discount}] \cdot p[\text{bd\_to\_il\_freq}] \cdot p[\text{bd\_transit\_coef}] \cdot x_{\text{bd}}^{\text{cCDCA}}\\
    &r_{\text{bd to il}}^{\text{cSBA}} &&= p[\text{bd\_discount}] \cdot p[\text{bd\_to\_il\_freq}] \cdot p[\text{bd\_transit\_coef}] \cdot x_{\text{bd}}^{\text{cSBA}}\\
    &r_{\text{il from bd}}^{\text{cCA}} &&= r_{\text{bd to il}}^{\text{cCA}}\\
    &r_{\text{il from bd}}^{\text{cCDCA}} &&= r_{\text{bd to il}}^{\text{cCDCA}}\\
    &r_{\text{il from bd}}^{\text{cSBA}} &&= r_{\text{bd to il}}^{\text{cSBA}}
\end{alignat*}
\paragraph{Ileum vs. Colon}
\begin{alignat*}{2}
    &r_{\text{il to co}}^{\text{cCA}} &&= p[\text{il\_to\_co\_freq}] \cdot p[\text{il\_transit\_coef}] \cdot x_{\text{il}}^{\text{cCA}}\\
    &r_{\text{il to co}}^{\text{cCDCA}} &&= p[\text{il\_to\_co\_freq}] \cdot p[\text{il\_transit\_coef}] \cdot x_{\text{il}}^{\text{cCDCA}}\\
    &r_{\text{il to co}}^{\text{cSBA}} &&= p[\text{il\_to\_co\_freq}] \cdot p[\text{il\_transit\_coef}] \cdot x_{\text{il}}^{\text{cSBA}}\\
    &r_{\text{il to co}}^{\text{uCA}} &&= p[\text{il\_to\_co\_freq}] \cdot p[\text{il\_transit\_coef}] \cdot x_{\text{il}}^{\text{uCA}}\\
    &r_{\text{il to co}}^{\text{uCDCA}} &&= p[\text{il\_to\_co\_freq}] \cdot p[\text{il\_transit\_coef}] \cdot x_{\text{il}}^{\text{uCDCA}}\\
    &r_{\text{il to co}}^{\text{uSBA}} &&= p[\text{il\_to\_co\_freq}] \cdot p[\text{il\_transit\_coef}] \cdot x_{\text{il}}^{\text{uSBA}}\\
    &r_{\text{co from il}}^{\text{cCA}} &&= r_{\text{il to co}}^{\text{cCA}}\\
    &r_{\text{co from il}}^{\text{cCDCA}} &&= r_{\text{il to co}}^{\text{cCDCA}}\\
    &r_{\text{co from il}}^{\text{cSBA}} &&= r_{\text{il to co}}^{\text{cSBA}}\\
    &r_{\text{co from il}}^{\text{uCA}} &&= r_{\text{il to co}}^{\text{uCA}}\\
    &r_{\text{co from il}}^{\text{uCDCA}} &&= r_{\text{il to co}}^{\text{uCDCA}}\\
    &r_{\text{co from il}}^{\text{uSBA}} &&= r_{\text{il to co}}^{\text{uSBA}}
\end{alignat*}
\paragraph{Plasma vs. Liver}
\begin{alignat*}{2}
    &r_{\text{pl to li}}^{\text{cCA}} &&= p[\text{hep\_ratio\_conj\_tri}] \cdot p[\text{pl\_flow\_thru\_li}] \cdot \frac{x_{\text{pl}}^{\text{cCA}}}{p[\text{pl\_volume}]}\\
    &r_{\text{pl to li}}^{\text{cCDCA}} &&= p[\text{hep\_ratio\_conj\_di}] \cdot p[\text{pl\_flow\_thru\_li}] \cdot \frac{x_{\text{pl}}^{\text{cCDCA}}}{p[\text{pl\_volume}]}\\
    &r_{\text{pl to li}}^{\text{cSBA}} &&= p[\text{hep\_ratio\_conj\_di}] \cdot p[\text{pl\_flow\_thru\_li}] \cdot \frac{x_{\text{pl}}^{\text{cSBA}}}{p[\text{pl\_volume}]}\\
    &r_{\text{pl to li}}^{\text{uCA}} &&= p[\text{hep\_ratio\_unconj\_tri}] \cdot p[\text{pl\_flow\_thru\_li}] \cdot \frac{x_{\text{pl}}^{\text{uCA}}}{p[\text{pl\_volume}]}\\
    &r_{\text{pl to li}}^{\text{uCDCA}} &&= p[\text{hep\_ratio\_unconj\_di}] \cdot p[\text{pl\_flow\_thru\_li}] \cdot \frac{x_{\text{pl}}^{\text{uCDCA}}}{p[\text{pl\_volume}]}\\
    &r_{\text{pl to li}}^{\text{uSBA}} &&= p[\text{hep\_ratio\_unconj\_di}] \cdot p[\text{pl\_flow\_thru\_li}] \cdot \frac{x_{\text{pl}}^{\text{uSBA}}}{p[\text{pl\_volume}]}\\
    &r_{\text{li from pl}}^{\text{cCA}} &&= r_{\text{pl to li}}^{\text{cCA}} + r_{\text{pl to li}}^{\text{uCA}}\\
    &r_{\text{li from pl}}^{\text{cCDCA}} &&= r_{\text{pl to li}}^{\text{cCDCA}} + r_{\text{pl to li}}^{\text{uCDCA}}\\
    &r_{\text{li from pl}}^{\text{cSBA}} &&= r_{\text{pl to li}}^{\text{cSBA}} + r_{\text{pl to li}}^{\text{uSBA}}
\end{alignat*}

\subsubsection{Deconjugation fluxes}
\begin{alignat*}{2}
    &r_{\text{il deconj}}^{\text{cCA}} &&= p[\text{gut\_deconj\_freq\_il}] \cdot x_{\text{il}}^{\text{cCA}}\\
    &r_{\text{il deconj}}^{\text{cCDCA}} &&= p[\text{gut\_deconj\_freq\_il}] \cdot x_{\text{il}}^{\text{cCDCA}}\\
    &r_{\text{il deconj}}^{\text{cSBA}} &&= p[\text{gut\_deconj\_freq\_il}] \cdot x_{\text{il}}^{\text{cSBA}}\\
    &r_{\text{co deconj}}^{\text{cCA}} &&= p[\text{gut\_deconj\_freq\_co}] \cdot x_{\text{co}}^{\text{cCA}}\\
    &r_{\text{co deconj}}^{\text{cCDCA}} &&= p[\text{gut\_deconj\_freq\_co}] \cdot x_{\text{co}}^{\text{cCDCA}}\\
    &r_{\text{co deconj}}^{\text{cSBA}} &&= p[\text{gut\_deconj\_freq\_co}] \cdot x_{\text{co}}^{\text{cSBA}}\\
    &r_{\text{il deconj}}^{\text{uCA}} &&= r_{\text{il deconj}}^{\text{cCA}}\\
    &r_{\text{il deconj}}^{\text{uCDCA}} &&= r_{\text{il deconj}}^{\text{cCDCA}}\\
    &r_{\text{il deconj}}^{\text{uSBA}} &&= r_{\text{il deconj}}^{\text{cSBA}}\\
    &r_{\text{co deconj}}^{\text{uCA}} &&= r_{\text{co deconj}}^{\text{cCA}}\\
    &r_{\text{co deconj}}^{\text{uCDCA}} &&= r_{\text{co deconj}}^{\text{cCDCA}}\\
    &r_{\text{co deconj}}^{\text{uSBA}} &&= r_{\text{co deconj}}^{\text{cSBA}}
\end{alignat*}

\subsubsection{Biotransformation fluxes}
\begin{alignat*}{2}
    &r_{\text{co biotr}}^{\text{uCA}} &&= p[\text{gut\_biotr\_freq\_CA}] \cdot x_{\text{co}}^{\text{uCA}}\\
    &r_{\text{co biotr}}^{\text{uCDCA}} &&= p[\text{gut\_biotr\_freq\_CA}] \cdot x_{\text{co}}^{\text{uCDCA}}\\
    &r_{\text{co biotr}}^{\text{uSBA}} &&= r_{\text{co biotr}}^{\text{uCA}} + r_{\text{co biotr}}^{\text{uCDCA}}
\end{alignat*}

\subsubsection{Active uptake fluxes}
\begin{alignat*}{2}
    &r_{\text{il au}}^{\text{cCA}} &&= p[\text{max\_asbt\_rate}] \cdot x_{\text{il}}^{\text{cCA}}\\
    &r_{\text{il au}}^{\text{cCDCA}} &&= p[\text{max\_asbt\_rate}] \cdot x_{\text{il}}^{\text{cCDCA}}\\
    &r_{\text{il au}}^{\text{cSBA}} &&= p[\text{max\_asbt\_rate}] \cdot x_{\text{il}}^{\text{cSBA}}\\
    &r_{\text{il au}}^{\text{uCA}} &&= p[\text{max\_asbt\_rate}] \cdot x_{\text{il}}^{\text{uCA}}\\
    &r_{\text{il au}}^{\text{uCDCA}} &&= p[\text{max\_asbt\_rate}] \cdot x_{\text{il}}^{\text{uCDCA}}\\
    &r_{\text{il au}}^{\text{uSBA}} &&= p[\text{max\_asbt\_rate}] \cdot x_{\text{il}}^{\text{uSBA}}
\end{alignat*}

\subsubsection{Passive uptake fluxes}
\begin{alignat*}{2}
    &r_{\text{il pu}}^{\text{uCA}} &&= p[\text{gut\_pu\_freq\_il}] \cdot x_{\text{il}}^{\text{uCA}}\\
    &r_{\text{il pu}}^{\text{uCDCA}} &&= p[\text{gut\_pu\_freq\_il}] \cdot x_{\text{il}}^{\text{uCDCA}}\\
    &r_{\text{il pu}}^{\text{uSBA}} &&= p[\text{gut\_pu\_freq\_il}] \cdot x_{\text{il}}^{\text{uSBA}}\\
    &r_{\text{co pu}}^{\text{uCA}} &&= p[\text{gut\_pu\_freq\_co}] \cdot x_{\text{co}}^{\text{uCA}}\\
    &r_{\text{co pu}}^{\text{uCDCA}} &&= p[\text{gut\_pu\_freq\_co}] \cdot x_{\text{co}}^{\text{uCDCA}}\\
    &r_{\text{co pu}}^{\text{uSBA}} &&= p[\text{gut\_pu\_freq\_co}] \cdot x_{\text{co}}^{\text{uSBA}}
\end{alignat*}

\subsubsection{Reabsorption fluxes}
\begin{alignat*}{2}
    &r_{\text{li from gut}}^{\text{cCA}} &&= p[\text{hep\_ratio\_conj\_tri}] \cdot r_{\text{il au}}^{\text{cCA}} + p[\text{hep\_ratio\_unconj\_tri}] \cdot (r_{\text{il au}}^{\text{uCA}} + r_{\text{il pu}}^{\text{uCA}} + r_{\text{co pu}}^{\text{uCA}})\\
    &r_{\text{li from gut}}^{\text{cCDCA}} &&= p[\text{hep\_ratio\_conj\_di}] \cdot r_{\text{il au}}^{\text{cCDCA}} + p[\text{hep\_ratio\_unconj\_di}] \cdot (r_{\text{il au}}^{\text{uCDCA}} + r_{\text{il pu}}^{\text{uCDCA}} + r_{\text{co pu}}^{\text{uCDCA}})\\
    &r_{\text{li from gut}}^{\text{cSBA}} &&= p[\text{hep\_ratio\_conj\_di}] \cdot r_{\text{il au}}^{\text{cSBA}} + p[\text{hep\_ratio\_unconj\_di}] \cdot (r_{\text{il au}}^{\text{uSBA}} + r_{\text{il pu}}^{\text{uSBA}} + r_{\text{co pu}}^{\text{uSBA}})\\
    &r_{\text{pl from gut}}^{\text{cCA}} &&= (1 - p[\text{hep\_ratio\_conj\_tri}]) \cdot r_{\text{il au}}^{\text{cCA}}\\
    &r_{\text{pl from gut}}^{\text{cCDCA}} &&= (1 - p[\text{hep\_ratio\_conj\_di}]) \cdot r_{\text{il au}}^{\text{cCDCA}}\\
    &r_{\text{pl from gut}}^{\text{cSBA}} &&= (1 - p[\text{hep\_ratio\_conj\_di}]) \cdot r_{\text{il au}}^{\text{cSBA}}\\
    &r_{\text{pl from gut}}^{\text{uCA}} &&= (1 - p[\text{hep\_ratio\_unconj\_tri}]) \cdot (r_{\text{il au}}^{\text{uCA}} + r_{\text{il pu}}^{\text{uCA}} + r_{\text{co pu}}^{\text{uCA}})\\
    &r_{\text{pl from gut}}^{\text{uCDCA}} &&= (1 - p[\text{hep\_ratio\_unconj\_di}]) \cdot (r_{\text{il au}}^{\text{uCDCA}} + r_{\text{il pu}}^{\text{uCDCA}} + r_{\text{co pu}}^{\text{uCDCA}})\\
    &r_{\text{pl from gut}}^{\text{uSBA}} &&= (1 - p[\text{hep\_ratio\_unconj\_di}]) \cdot (r_{\text{il au}}^{\text{uSBA}} + r_{\text{il pu}}^{\text{uSBA}} + r_{\text{co pu}}^{\text{uSBA}})
\end{alignat*}

\subsubsection{Excretion fluxes}
\begin{alignat*}{2}
    &r_{\text{co to fe}}^{\text{cCA}} &&= p[\text{co\_to\_fe\_freq}] \cdot p[\text{co\_transit\_coef}] \cdot x_{\text{co}}^{\text{cCA}}\\
    &r_{\text{co to fe}}^{\text{cCDCA}} &&= p[\text{co\_to\_fe\_freq}] \cdot p[\text{co\_transit\_coef}] \cdot x_{\text{co}}^{\text{cCDCA}}\\
    &r_{\text{co to fe}}^{\text{cSBA}} &&= p[\text{co\_to\_fe\_freq}] \cdot p[\text{co\_transit\_coef}] \cdot x_{\text{co}}^{\text{cSBA}}\\
    &r_{\text{co to fe}}^{\text{uCA}} &&= p[\text{co\_to\_fe\_freq}] \cdot p[\text{co\_transit\_coef}] \cdot x_{\text{co}}^{\text{uCA}}\\
    &r_{\text{co to fe}}^{\text{uCDCA}} &&= p[\text{co\_to\_fe\_freq}] \cdot p[\text{co\_transit\_coef}] \cdot x_{\text{co}}^{\text{uCDCA}}\\
    &r_{\text{co to fe}}^{\text{uSBA}} &&= p[\text{co\_to\_fe\_freq}] \cdot p[\text{co\_transit\_coef}] \cdot x_{\text{co}}^{\text{uSBA}}\\
    &r_{\text{fe from co}}^{\text{cCA}} &&= r_{\text{co to fe}}^{\text{cCA}}\\
    &r_{\text{fe from co}}^{\text{cCDCA}} &&= r_{\text{co to fe}}^{\text{cCDCA}}\\
    &r_{\text{fe from co}}^{\text{cSBA}} &&= r_{\text{co to fe}}^{\text{cSBA}}\\
    &r_{\text{fe from co}}^{\text{uCA}} &&= r_{\text{co to fe}}^{\text{uCA}}\\
    &r_{\text{fe from co}}^{\text{uCDCA}} &&= r_{\text{co to fe}}^{\text{uCDCA}}\\
    &r_{\text{fe from co}}^{\text{uSBA}} &&= r_{\text{co to fe}}^{\text{uSBA}}
\end{alignat*}

\subsubsection{Sulfation fluxes}
In human, a portion of secondary bile acids exists in sulfated form (\citet{ba_sulfation}). However, according to \citet{sips_ba_model}, the process of desulfation happens at a very slow rate. This means only a small fraction of sulfated bile acids can re-enter the enterohepatic circulation through desulfation, while the majority is excreted through feces. Hence, for simplicity, we account for sulfation by modeling its end outcome: the additional excretion of secondary bile acids into feces, which is again modeled with first-order dynamics.
\begin{alignat*}{2}
    &r_{\text{co sulfate}}^{\text{cSBA}} &&= p[\text{co\_sulfate\_freq}] \cdot p[\text{co\_transit\_coef}] \cdot x_{\text{co}}^{\text{cSBA}}\\
    &r_{\text{co sulfate}}^{\text{uSBA}} &&= p[\text{co\_sulfate\_freq}] \cdot p[\text{co\_transit\_coef}] \cdot x_{\text{co}}^{\text{uSBA}}\\
    &r_{\text{fe sulfate}}^{\text{cSBA}} &&= r_{\text{co sulfate}}^{\text{cSBA}}\\
    &r_{\text{fe sulfate}}^{\text{uSBA}} &&= r_{\text{co sulfate}}^{\text{uSBA}}
\end{alignat*}
\clearpage
    \end{strip}
\endgroup

\section{Reinforcement Learning Specifications And Training Statistics}\label{apd:rl_setup}
\subsection{Additional Details of RL Setup}
\paragraph{State:} the state vector comprises bile acid levels $\mathbi{x}_{bile\ acids}$ and parameter values that represent host enzyme activity levels $\mathbi{p}_{adapt}$. Specifically, $\mathbi{x}_{bile\ acids}$ includes 30 variables across six bile acid species and six organs, denoted by $\{x_{OG}^{BA}, \forall BA \in\{\text{cCA, cCDCA, cSBA}\}, OG \in\{\text{li, bd, il, co, pl, fe}\}\}$ and $\{x_{OG}^{BA}, \forall BA \in\{\text{uCA, uCDCA, uSBA}\}, OG \in\{\text{il, co, pl, fe}\}\}$. Unconjugated bile acids in the liver and bile duct are not included as we assume bile acids are always conjugated in these two organs. $\mathbi{p}_{adapt}$ includes $\{p[\text{synthesis}]\mathrel{,}p[\text{syn\_frac\_CA}]\mathrel{,}p[\text{hep\_ratio\_conj\_tri}]\mathrel{,}p[\text{hep\_ratio\_conj\_di}]\mathrel{,}p[\text{max\_asbt\_rate}]\}$, same as the parameters in the action space.

\paragraph{Reward:}
Here we introduce the additional reward terms used to ensure that the RL agent generates bile acid levels that are physiologically plausible while keeping the regulatable parameters close to their values in healthy conditions:
\subparagraph{Enforcing physiologically plausible ranges.} All bile acid levels are nonnegative, and they are upper bound by the maximum bile acid storage capacity in each organ. This reward term penalizes the RL agent for violating physiologically plausible bile acids ranges, pre-specified according to the literature. If the agent takes actions that lead to bile acid levels outside of any of the ranges, a large negative reward (VIOLATE\_BOUNDARY\_PENALTY) is applied, and the episode ends. We set VIOLATE\_BOUNDARY\_PENALTY to be -100.
\subparagraph{Reward for not violating boundary conditions.} To encourage the RL agent to take actions that lead to physiologically plausible states, we design this auxiliary reward term to give a small constant positive reward for every RL step that stays within the physiologically plausible ranges. We set the small positive reward to be 4.
\subparagraph{Encouraging parameters resembling healthy conditions.} This reward term encourages the RL agent to keep the regulatable parameters close to their values in healthy conditions. A negative reward is applied to penalize deviation of parameters from healthy conditions. We quantify the deviation as the smallest number of steps it takes to move from the healthy condition values to current values. For example, for $p[\text{synthesis}]$, a 25\% higher or lower fold change is allowed at every step, then a value that is 156.25\% of the original value takes at least two steps to reach, resulting in a deviation of 2. We sum up the deviation across all parameters and set the negative value of the sum as the reward. We multiply the negative deviation sum by a coefficient within $[0,1]$ to match the range of other reward terms. Roughly each parameter is allowed a deviation penalty of at most 10 (and we clip the penalty at 10 if the deviation exceeds 10), and five adaptable parameters result in a total penalty of at most 50. We set the coefficient of this reward term to be 0.02 such that the penalty for parameter deviation is at most 1.

We also specify additional implementation details of the reward terms mentioned in the main text:
\subparagraph{Minimizing toxicity.} The maximum possible $LE$ is calculated as the upper bound of physiologically plausible bile acid in the liver integrated over one day.
\subparagraph{Resembling real-world patient data.} Bile duct ligation experiments in mice showed that the adaptation usually stabilized after the first two weeks. Because ligation is an extreme model of bile duct obstruction, we allowed a four-week adaptation period in the RL simulation before introducing the fitting error to encourage RL trajectories that are similar to stabilized patient data. Because bile acid measurements are roughly log-normally distributed, log transform is applied to both the actual measurements and the RL-generated bile acid states. We set the cap for fitting error, CAP, to be a generous value of 20. We set the coefficient $\lambda_{\text{error}}$ to be 0.2 such that the fitting error is bounded at 4, giving more emphasis on the data-guided fitting error than the other biology-inspired reward terms that are at most 1. 

\subsection{RL algorithm training}
The actor and critic in PPO shared a neural network with the architecture of a three-layer multilayer perceptron with 100, 50, and 25 hidden units in each layer. We set the learning rate to be 0.002. We normalized the states during the RL training for easier optimization of the algorithm. We used 16 vectorized environments on different processes to speed up training. We leave the other training parameters to be the default of the implementation, including a discount factor of 0.99 to mimic the optimization of long-term homeostasis. We trained the RL agent for 4,000,000 environment steps to ensure convergence of the algorithm. 

\subsection{RL training statistics}
\figureref{fig:rl_step_vs_reward} shows the RL training statistics. We see that PPO is able to converge across different values of $p[\text{bd\_max\_flow}]$, with smaller $p[\text{bd\_max\_flow}]$, i.e., more severe obstruction of the bile ducts, resulting in less reward. This is expected because with more severe diseases, balancing between minimizing liver toxicity and maintaining cholesterol elimination and digestion becomes more difficult, and restoring homeostasis requires a higher level of adaptation that will result in larger deviation from healthy conditions. However, after training completes, as the simulation in \figureref{fig:rl_day_vs_fitting_error} shows, after the initial adaptation period right after the introduction of PSC pathophysiology, simulation corresponding to $p[\text{bd\_max\_flow}]$=3 $\mu$mol/min generates fasting plasma bile acid profiles closest to a representative PSC patient, shown by the smallest fitting error indicated by the green curve. This scenario represents a relatively severe bile duct obstruction. We chose to focus on this scenario for the later analysis under the assumption that a small fitting error from actual data implies that this is more likely to reflect what actually happens in PSC patients.

\begin{figure*}[htbp]
\floatconts
  {fig:rl_step_vs_reward}
  {\caption{PPO training steps vs. mean reward evaluated on 5 episodes, each curve represents a different level of disease severity reflected by $p[\text{bd\_max\_flow}]$.}}
  {\includegraphics[width=0.9\linewidth]{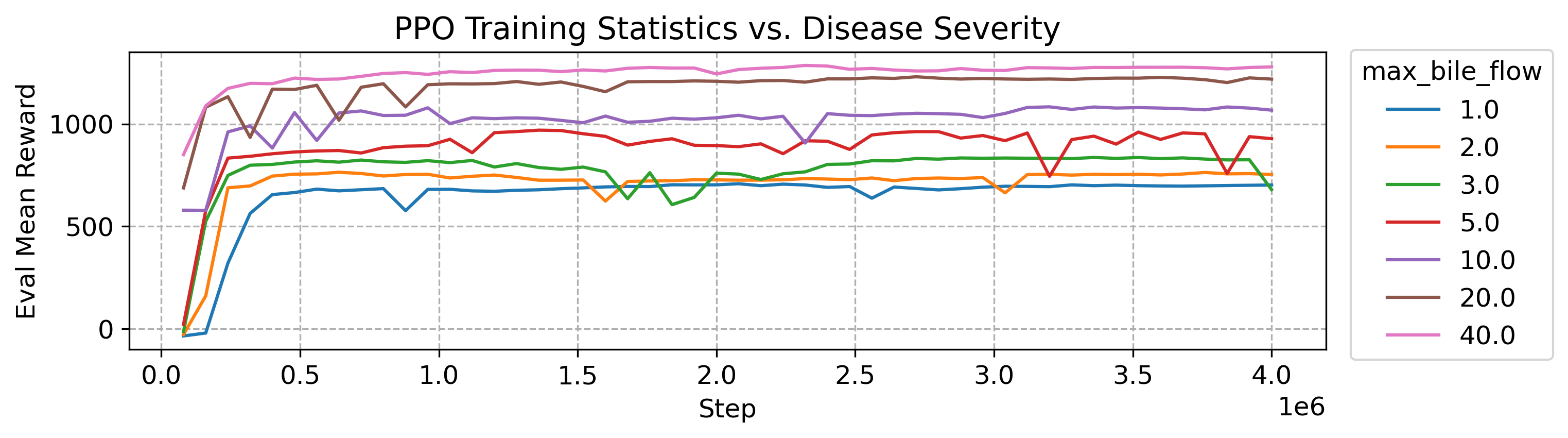}}
\end{figure*}

\begin{figure*}[htbp]
\floatconts
  {fig:rl_day_vs_fitting_error}
  {\caption{Fitting error between simulated fasting plasma bile acids (with adaptation) and fasting plasma bile acid measurement from a representative PSC patient. Bile acid measurement from the actual patient is cross-sectional, while the fitting error is calculated daily, over a 240-day simulation after introducing PSC pathophysiology. Each curve represents a different level of disease severity reflected by $p[\text{bd\_max\_flow}]$. The adaptive ODE parameters are obtained from PPO agents trained with the corresponding $p[\text{bd\_max\_flow}]$ for 4M steps.}}
  {\includegraphics[width=0.9\linewidth]{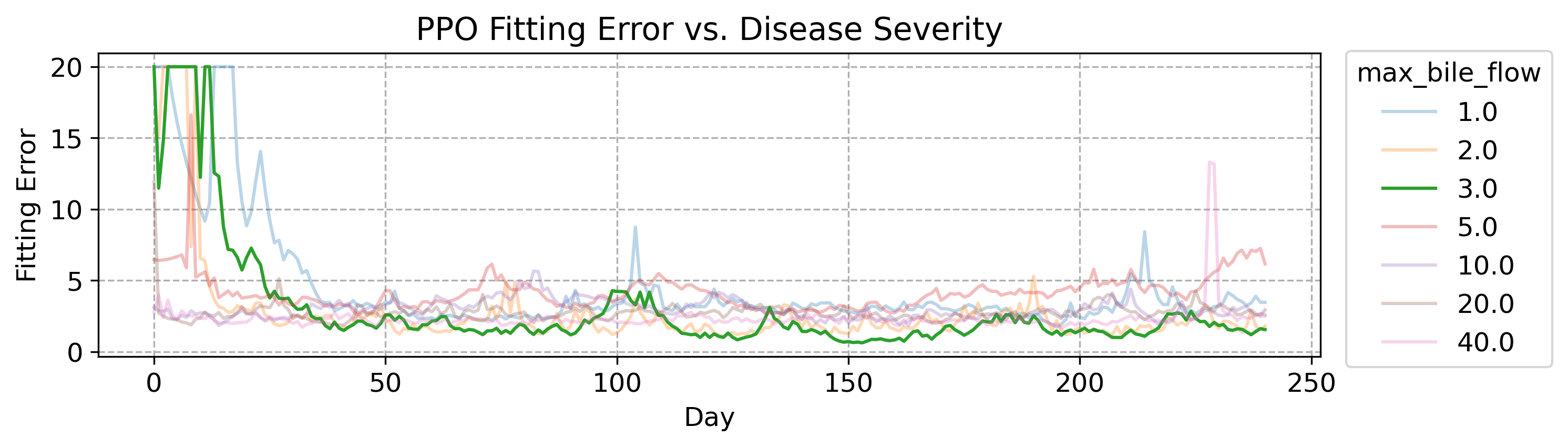}}
\end{figure*}

\clearpage
\section{Complete Bile Acid Dynamics After Introducing PSC Pathophysiology}\label{apd:complete_ba}
For all simulations in this section, PSC pathophysiology was simulated by setting $p[\text{bd\_max\_flow}]$=3 $\mu$mol/min at day 0, much lower than the maximum bile duct to the small intestine flux of 75 $\mu$mol/min observed in the healthy individual simulation. We initialized the bile acid values with their steady-state values under healthy conditions. This is to mimic one animal model of PSC --- where bile duct ligation surgeries are performed on healthy mice and PSC-like symptoms will develop in the mice. Other ODE parameters were also initialized using their values under healthy conditions. Only fasting values (8 AM) were plotted in the figures.

\subsection{Bile acid dynamics without RL}
\begin{figure*}[htbp]
\floatconts
  {fig:psc_ba_wo_rl}
  {\caption{Simulated 60-day bile acid dynamics after introducing PSC pathophysiology, without RL adaptations.}}
  {\includegraphics[width=1\linewidth]{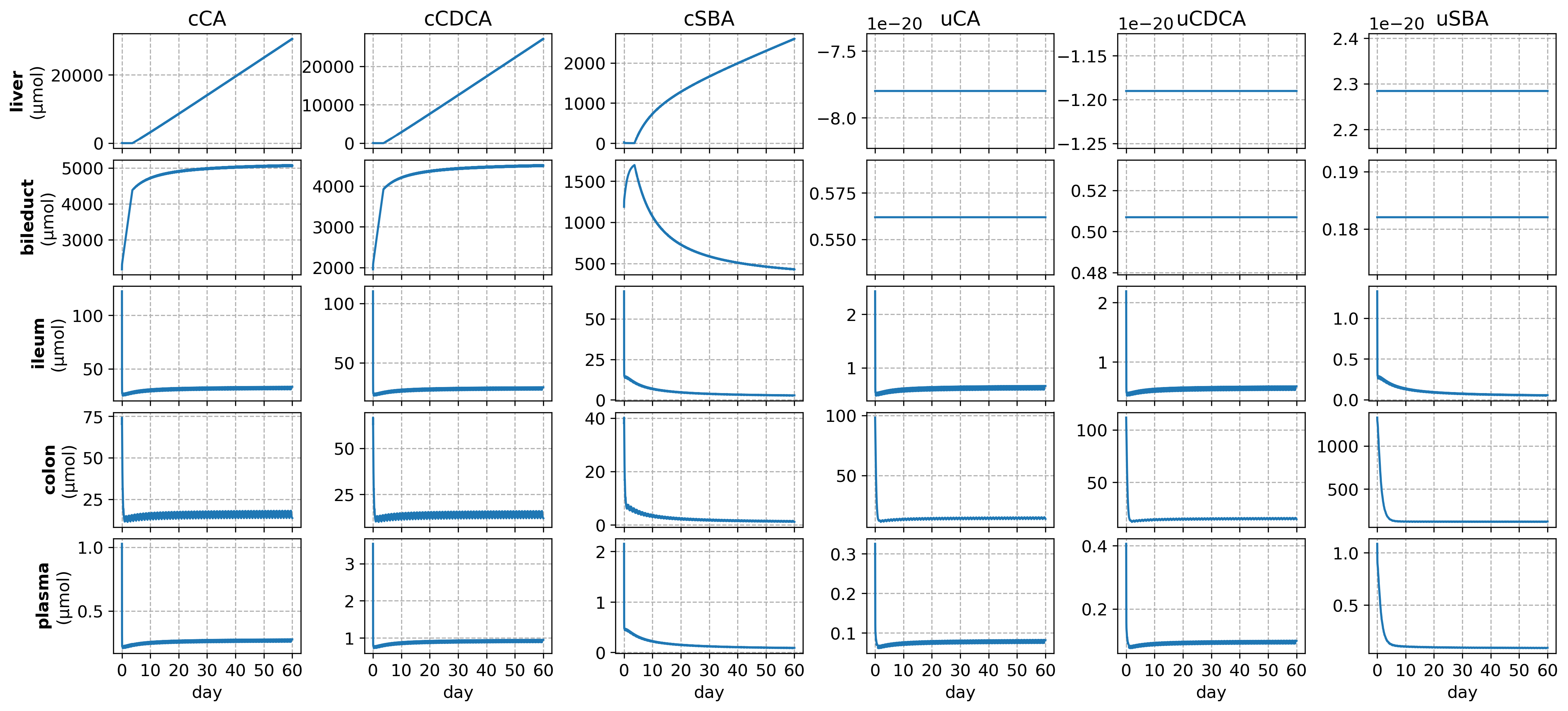}}
\end{figure*}

To see the simulated PSC bile acid dynamics without RL, we ran the simulation for 60 days after introducing PSC pathophysiology. We chose a simulation period of 60 days to be longer than the two-week adaptation period in the mouse experiment (after which the mouse liver adaptation stabilizes) and to achieve stable bile acid states (all organs other than the liver) in the simulation. Simulated meals at 8 AM, 2 PM, and 8 PM are included as usual. \figureref{fig:psc_ba_wo_rl} shows the complete simulated 60-day bile acid dynamics without RL adaptation. We observe unrealistically high liver bile acid levels and decreased plasma conjugated bile acids that contradict clinical data. Changing $p[\text{bd\_max\_flow}]$ to 10 $\mu$mol/min results in similar trends. Moreover, continuing the simulation would eventually run into numerical issues due to the ever-increasing liver bile acid levels.

These discrepancies arise from the flawed assumption that bile acid metabolism parameters are constants. While this assumption might be valid under healthy conditions, where a homeostatic state is maintained, it fails to account for the dynamic nature of PSC. Bile duct ligation experiments in mice have shown bile duct obstruction rapidly disrupts bile acid homeostasis, as reflected by the immediate shift in bile acid profiles upon introducing PSC pathophysiology into our model. When such disruption of homeostasis occurs, it is crucial to recognize that biological systems are inherently dynamic and have the ability to self-adapt in order to restore homeostasis. 

For instance, in cholestatic conditions, bile acids accumulate in the liver. The liver responds by suppressing the synthesis of bile acids to mitigate potential toxicity associated with excessive bile acid exposure. However, it is important to note that complete suppression of bile acid synthesis is not a feasible option. Bile acid synthesis is the sole source of incoming bile acids in the enterohepatic circulation. If synthesis were completely halted, it would eventually deplete bile acids in all compartments, thereby leaving no bile acids for the digestion of fat. Therefore, modeling bile acid metabolism under PSC pathophysiology requires accounting for the dynamic nature of the system, its self-adjusting capabilities, and the various mechanisms aimed at preserving homeostasis. We achieve this in our model by allowing the values of the biomolecular parameters to undergo continuous changes that ensure the preservation of homeostasis.

\subsection{Bile acid dynamics with RL (Patient 1)}
\begin{figure*}[htbp]
\floatconts
  {fig:psc_ba_w_rl}
  {\caption{Simulated 240-day bile acid dynamics trajectory after introducing PSC pathophysiology with adaptive ODE parameters obtained from a trained PPO agent.}}
  {\includegraphics[width=1\linewidth]{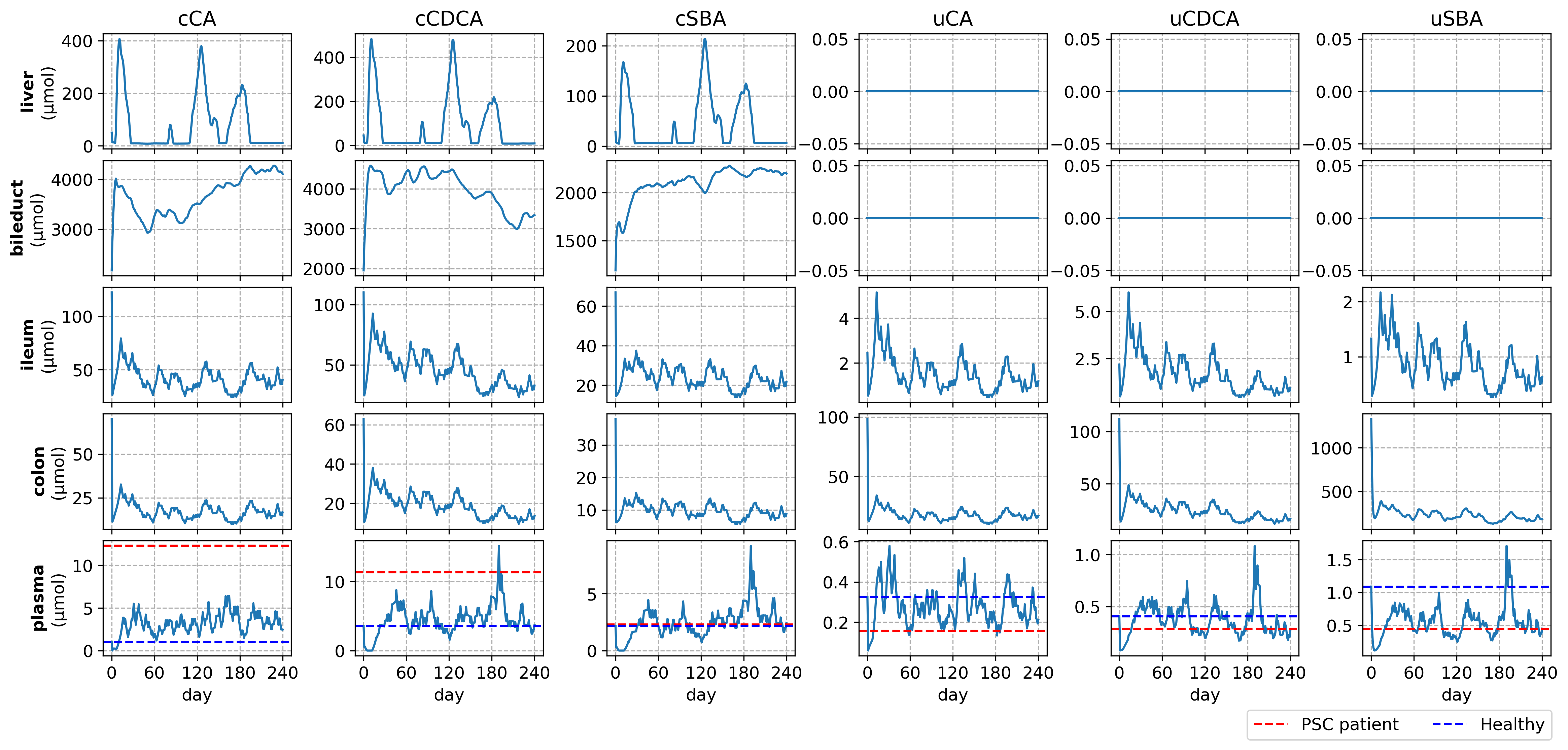}}
\end{figure*}

\figureref{fig:psc_ba_w_rl} shows the complete 240-day PSC bile acid dynamics with adaptive ODE parameters obtained from the trained PPO agent. As indicated by the red and blue dashed lines in the trajectories of plasma bile acids, the RL adaptation led to the elevation of plasma conjugated bile acids values and the decrease of plasma unconjugated bile acids. The resulting plasma bile acids profile was closer to the actual PSC patient measurements, correcting the artifacts in the simulation without RL where plasma bile acids decreased across all species.

\subsection{Bile Acid Dynamics with RL for ten random seeds (Patient 1)}
\begin{figure*}[htbp]
\floatconts
  {fig:psc_ba_w_rl_10_seeds}
  {\caption{Simulated 240-day bile acid dynamics trajectories after introducing PSC pathophysiology with adaptive ODE parameters obtained from a trained PPO agent and ten random seeds.}}
  {\includegraphics[width=1\linewidth]{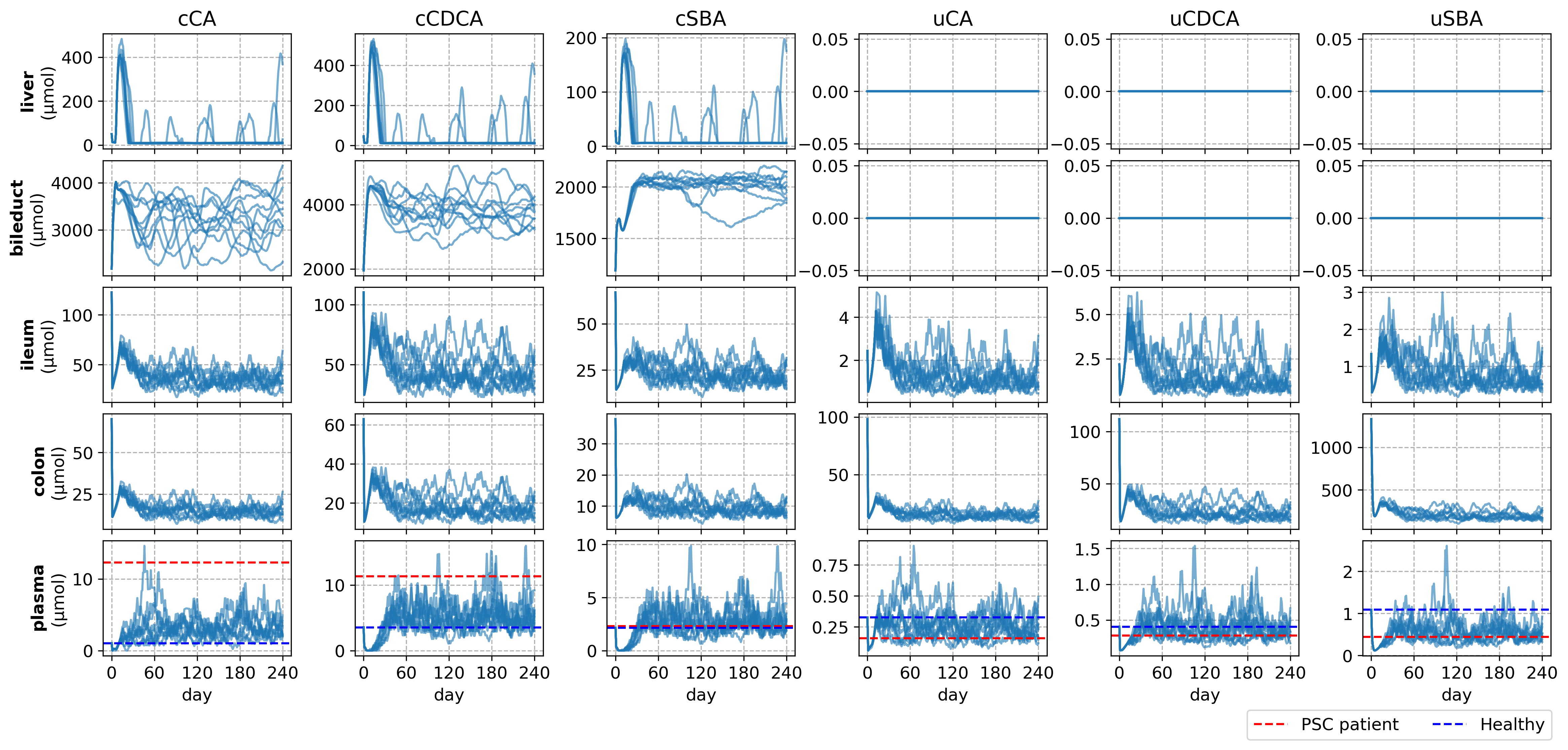}}
\end{figure*}

To understand how the stochasticity in the RL agent's actions affects the bile acid dynamics. \figureref{fig:psc_ba_w_rl_10_seeds} showed the 240-day PSC bile acid dynamics with adaptive ODE parameters obtained from the trained PPO agent for ten random seeds. Trajectories from all ten seeds shared a similar adaptation period in the beginning, where bile duct bile acids increased, ileum and colon bile acids decreased, and liver bile acids experienced a sharp increase and then decreased following the saturation of bile duct bile acids. After the initial adaptation period, trajectories from different seeds exhibited larger but bounded variability from the stochasticity of the RL actions. Interestingly, all trajectories contained sporadic peaks of liver bile acids, potentially explaining the episodic symptom flare-ups in PSC patients.

\subsection{\textit{In silico} evaluation of bile acid therapies (Patient 1)}
\begin{figure*}[htbp]
\floatconts
  {fig:complete_reduce_asbt}
  {\caption{Simulated 240-day bile acid dynamics after introducing PSC pathology with adaptive ODE parameters obtained from a trained PPO agent. Complete reduction of active uptake was simulated by forcing $p[\text{max\_asbt\_rate}]$ to be down-regulated unless it reaches the lower bound of its predefined physiological ranges.}}
  {\includegraphics[width=1\linewidth]{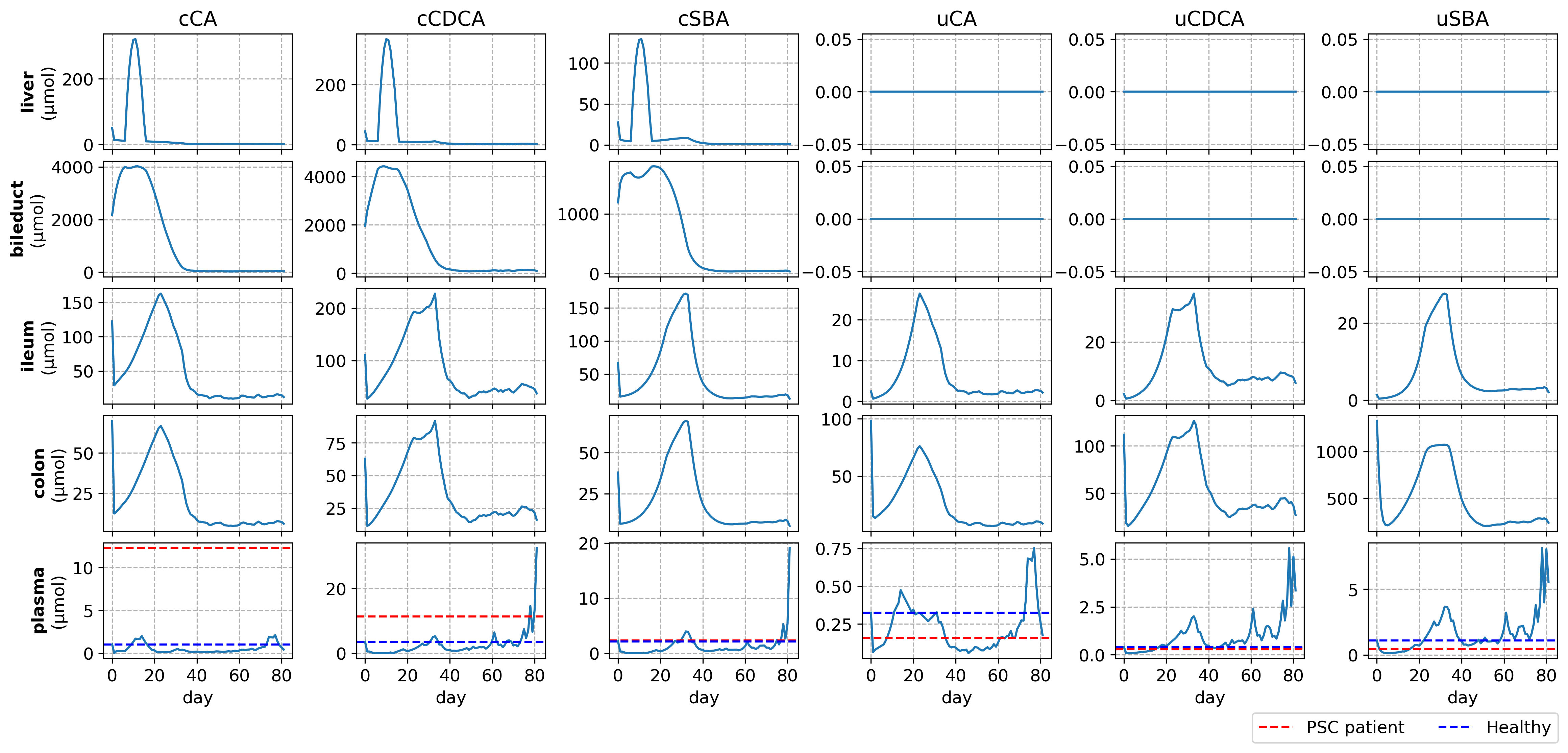}}
\end{figure*}

ASBT inhibitors such as maralixibat have been in clinical trials for PSC. As mentioned before, ASBT mediates the active uptake of bile acid from the ileum. We used our trained RL agent as an in-silico testbed to test ASBT inhibitors for PSC by decreasing the corresponding parameter $p[\text{max\_asbt\_rate}]$. \figureref{fig:complete_reduce_asbt} showed the resulting simulated trajectories. Compared to the case of no inhibition, the complete reduction of active uptake prevented peaks in liver bile acids after the initial adaptation period. However, significant increases in intestine and plasma bile acid levels led to premature termination of the simulation, rendering complete reduction a perhaps unrealistic strategy.

\begin{figure*}[htbp]
\floatconts
  {fig:half_reduce_asbt}
  {\caption{Simulated 240-day bile acid dynamics after introducing PSC pathology with adaptive ODE parameters obtained from a trained PPO agent and ten pre-specified random seeds. 50\% reduction of active uptake was simulated by forcing $p[\text{max\_asbt\_rate}]$ to be down-regulated unless it reaches or is below half of the lower bound of its predefined physiological ranges.}}
  {\includegraphics[width=1\linewidth]{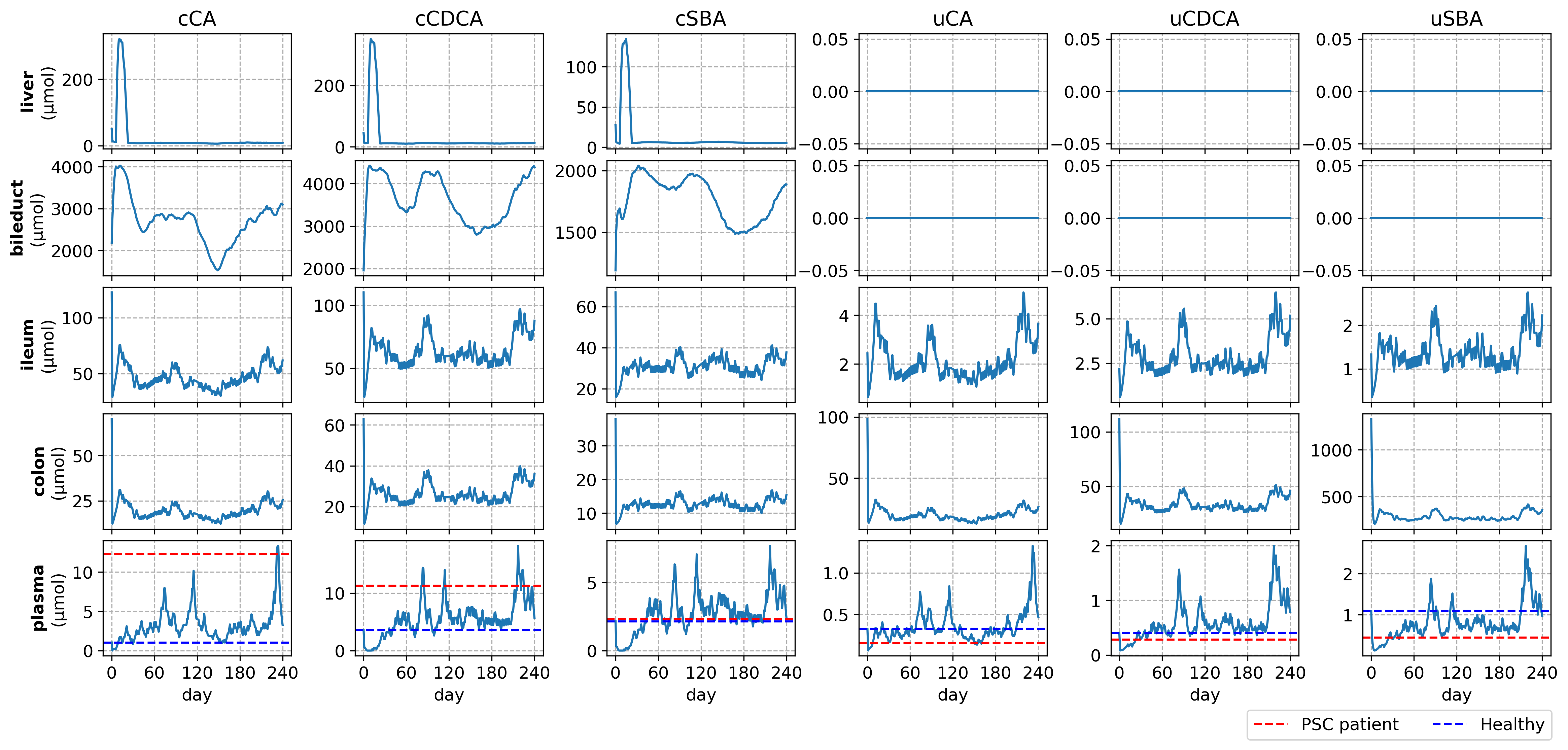}}
\end{figure*}

We further tested a more moderate strategy --- 50\% reduction of active uptake with our trained RL agent. \figureref{fig:half_reduce_asbt} showed a reduction of liver bile acid peaks compared to the case of no intervention, suggesting that partial suppression of active uptake might help alleviate the toxicity burden on the liver. Furthermore, unlike complete reduction of active uptake, all bile acid species were able to stay within plausible ranges across the organs, making partial reduction a more realistic strategy.

\clearpage
\begingroup
    \begin{strip}
\section{Bile Acid Dynamics After Introducing PSC Pathophysiology (Patient 2--4)}\label{apd:complete_ba_other_patients}
To avoid potential bias from patient selection, we replicated our experiments with four other representative patients who had the second to the fourth smallest sum of distances to all patients. We showed their bile acid dynamics as well as parameter adaptation trajectories below. For consistency with results in the main text, we chose the case of $p[\text{bd\_max\_flow}]$=3 $\mu$mol/min to train REMEDI across all cases.

In \tableref{tab:pl_ba_difference_five_patients}, we also showed the fitting error between REMEDI prediction trained with all patients and the plasma measurements from the representative PSC patient chosen in the main text.
    \end{strip}
\endgroup

\begin{table*}[hb]
\floatconts
  {tab:pl_ba_difference_five_patients}
  {\caption{Fitting Error ($\mu$mol, day 50 -- 60 averaged): REMEDI $-$ Data from PSC \textbf{Patient 1}.}}
  {\begin{tabular}{lllllll}
  \toprule
  \bfseries Scenario/Plasma & \bfseries cCA & \bfseries cCDCA & \bfseries cSBA & \bfseries uCA & \bfseries uCDCA & \bfseries uSBA\\
  \midrule
    REMEDI (without RL)  & -12.01  & -10.37  & -2.21  & -0.08  & -0.18  & -0.35\\
    REMEDI (trained with \textbf{patient 1})  & -9.79  & -5.27  & 0.67  & 0.05  & 0.16  & 0.15\\
    REMEDI (trained with \textbf{patient 2})  & -10.67  & -4.75  & 0.54  & 0.02  & 0.33  & 0.26\\
    REMEDI (trained with \textbf{patient 3})  & -9.56  & -7.32  & 0.19  & 0.15  & 0.10  & 0.22\\
    REMEDI (trained with \textbf{patient 4})  & -10.34  & -5.14  & 1.17  & 0.01  & 0.09  & 0.14\\
    REMEDI (trained with \textbf{patient 5})  & -9.06  & -9.51  & -0.88  & 0.14  & -0.12  & -0.07\\
  \bottomrule
  \end{tabular}}
\end{table*}

\begin{figure*}[htbp]
\floatconts
  {fig:psc_ba_w_rl_patient_2}
  {\caption{\textbf{Patient 2:} Simulated 240-day bile acid dynamics trajectory after introducing PSC pathophysiology with adaptive ODE parameters obtained from a trained PPO agent.}}
  {\includegraphics[width=1\linewidth]{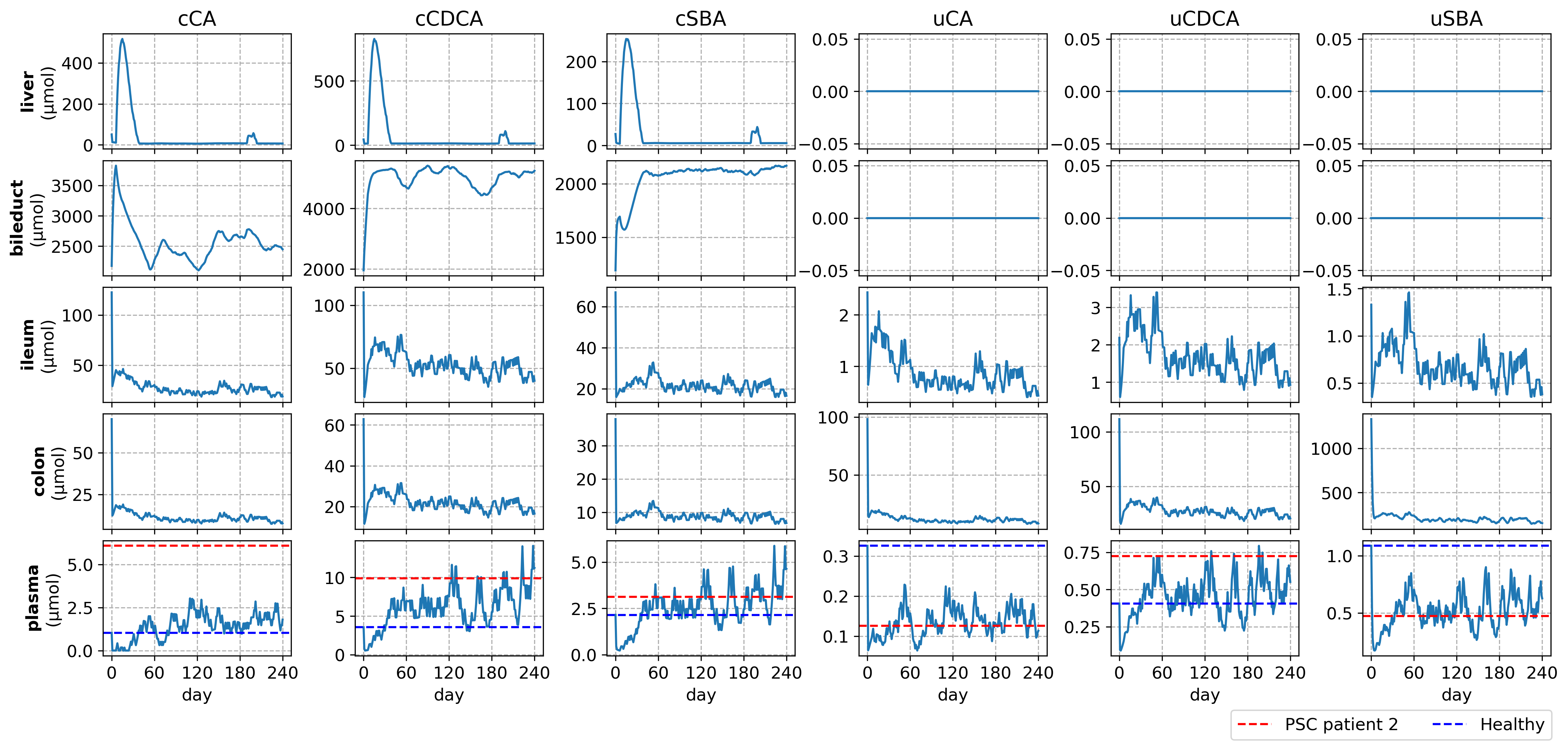}}
\end{figure*}

\begin{figure*}[htbp]
\floatconts
  {fig:psc_ode_param_w_rl_patient_2}
  {\caption{Adaptation of ODE parameters corresponding to the simulation scenario illustrated in \figureref{fig:psc_ba_w_rl_patient_2}.}}
  {\includegraphics[width=0.75\linewidth]{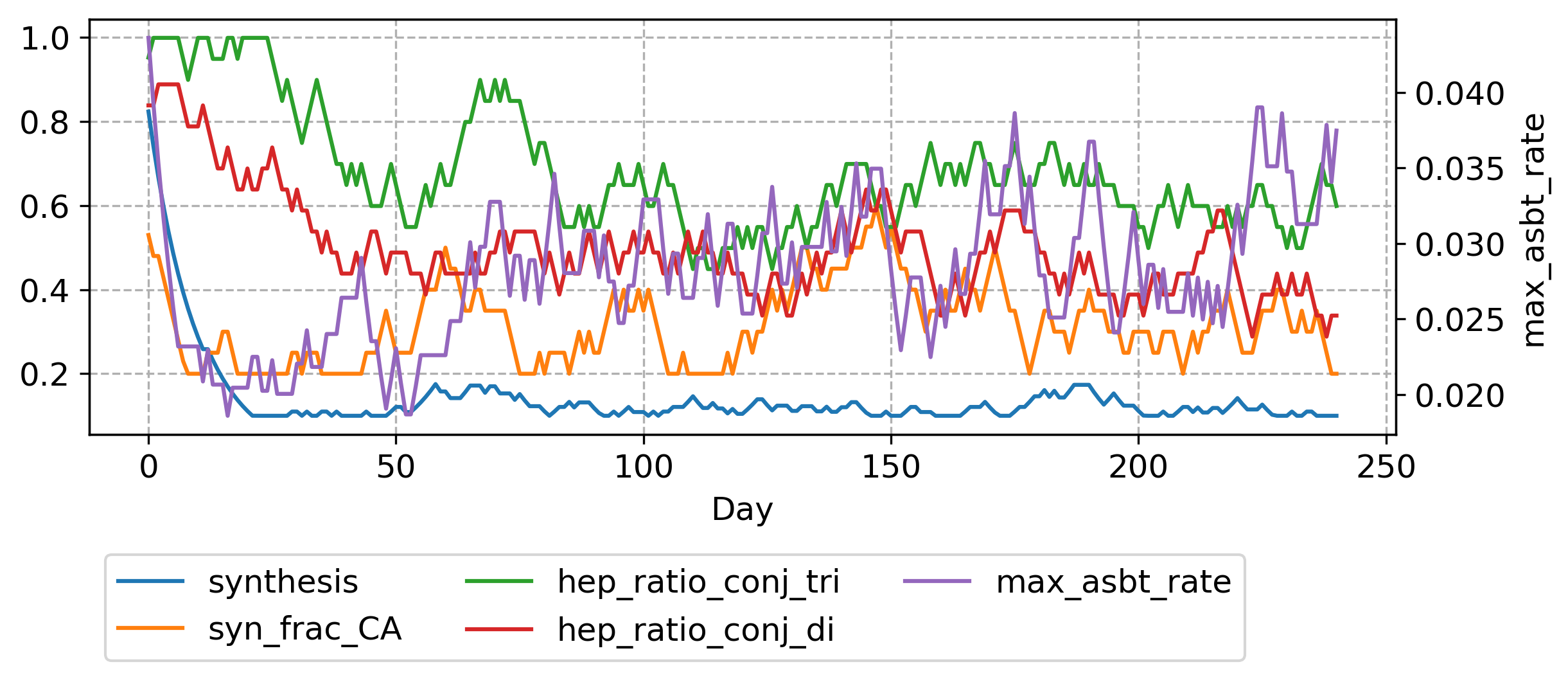}}
\end{figure*}

\begin{figure*}[htbp]
\floatconts
  {fig:psc_ba_w_rl_patient_3}
  {\caption{\textbf{Patient 3:} Simulated 240-day bile acid dynamics trajectory after introducing PSC pathophysiology with adaptive ODE parameters obtained from a trained PPO agent.}}
  {\includegraphics[width=1\linewidth]{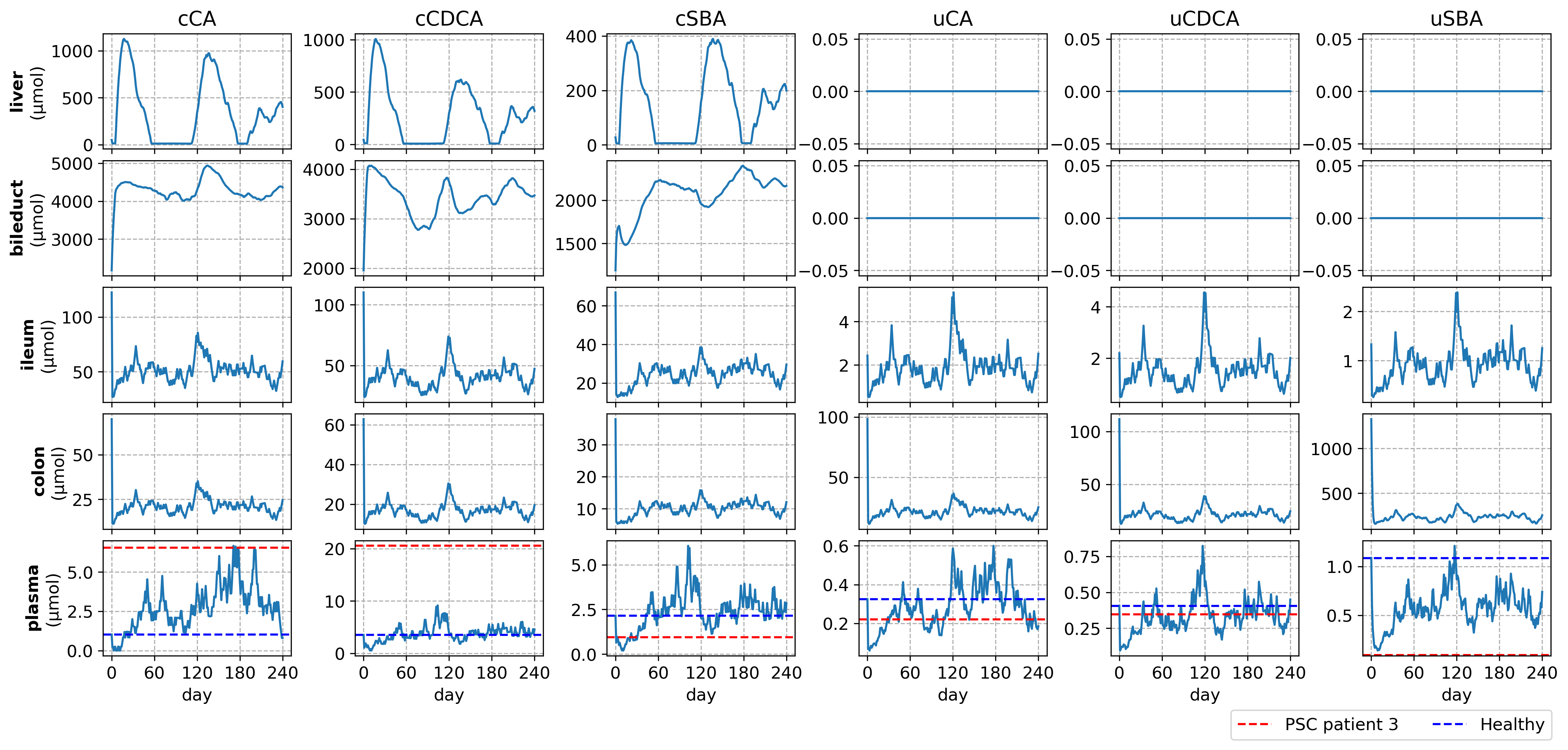}}
\end{figure*}

\begin{figure*}[htbp]
\floatconts
  {fig:psc_ode_param_w_rl_patient_3}
  {\caption{Adaptation of ODE parameters corresponding to the simulation scenario illustrated in \figureref{fig:psc_ba_w_rl_patient_3}.}}
  {\includegraphics[width=0.75\linewidth]{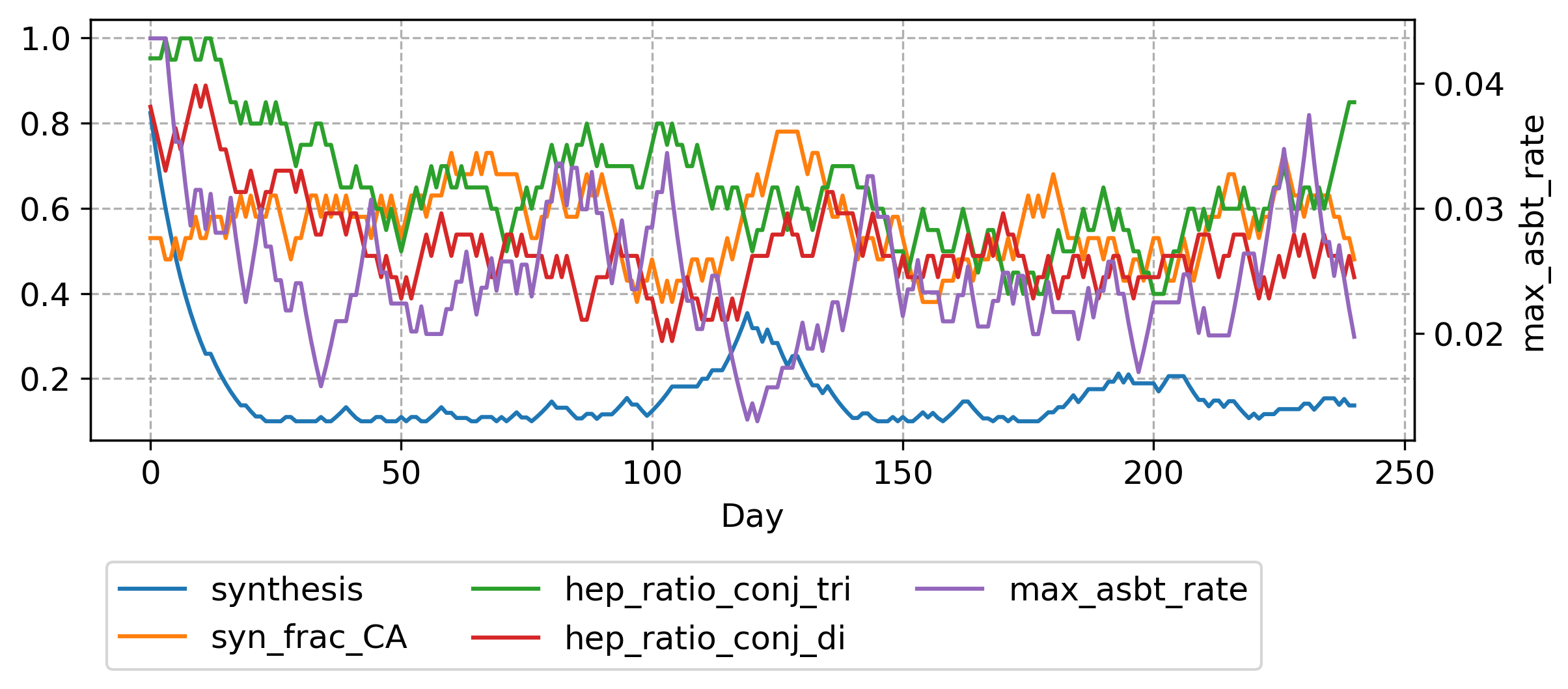}}
\end{figure*}

\begin{figure*}[htbp]
\floatconts
  {fig:psc_ba_w_rl_patient_4}
  {\caption{\textbf{Patient 4:} Simulated 240-day bile acid dynamics trajectory after introducing PSC pathophysiology with adaptive ODE parameters obtained from a trained PPO agent.}}
  {\includegraphics[width=1\linewidth]{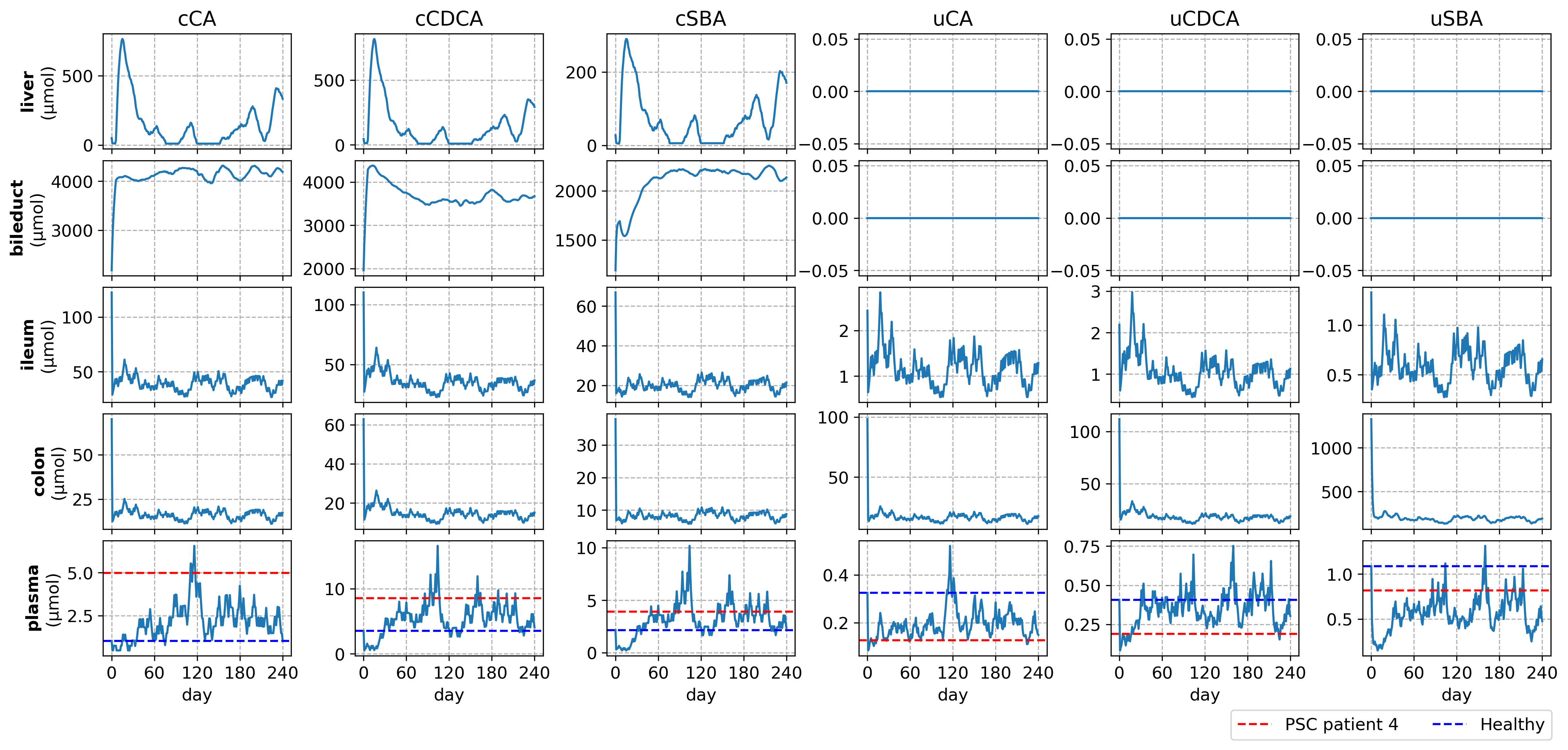}}
\end{figure*}

\begin{figure*}[htbp]
\floatconts
  {fig:psc_ode_param_w_rl_patient_4}
  {\caption{Adaptation of ODE parameters corresponding to the simulation scenario illustrated in \figureref{fig:psc_ba_w_rl_patient_4}.}}
  {\includegraphics[width=0.75\linewidth]{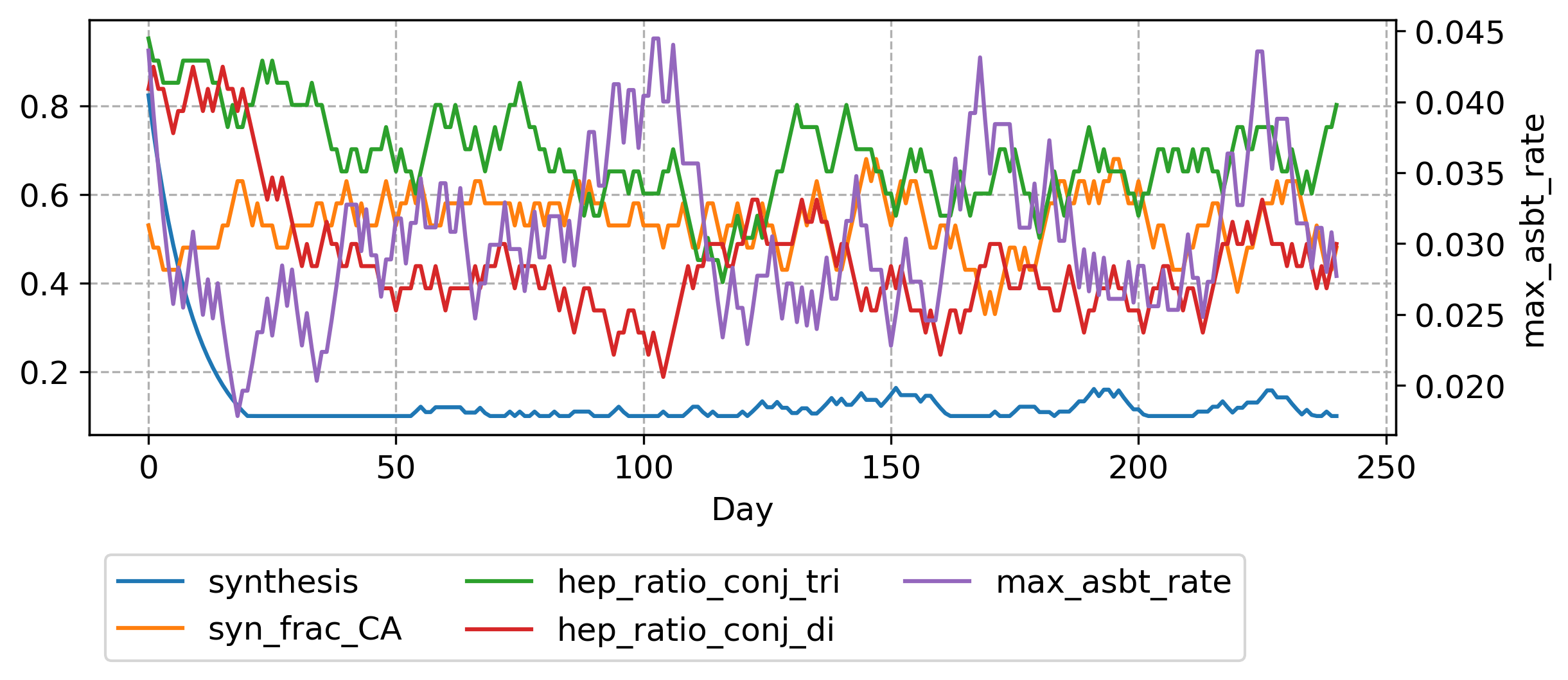}}
\end{figure*}

\begin{figure*}[htbp]
\floatconts
  {fig:psc_ba_w_rl_patient_5}
  {\caption{\textbf{Patient 5:} Simulated 240-day bile acid dynamics trajectory after introducing PSC pathophysiology with adaptive ODE parameters obtained from a trained PPO agent.}}
  {\includegraphics[width=1\linewidth]{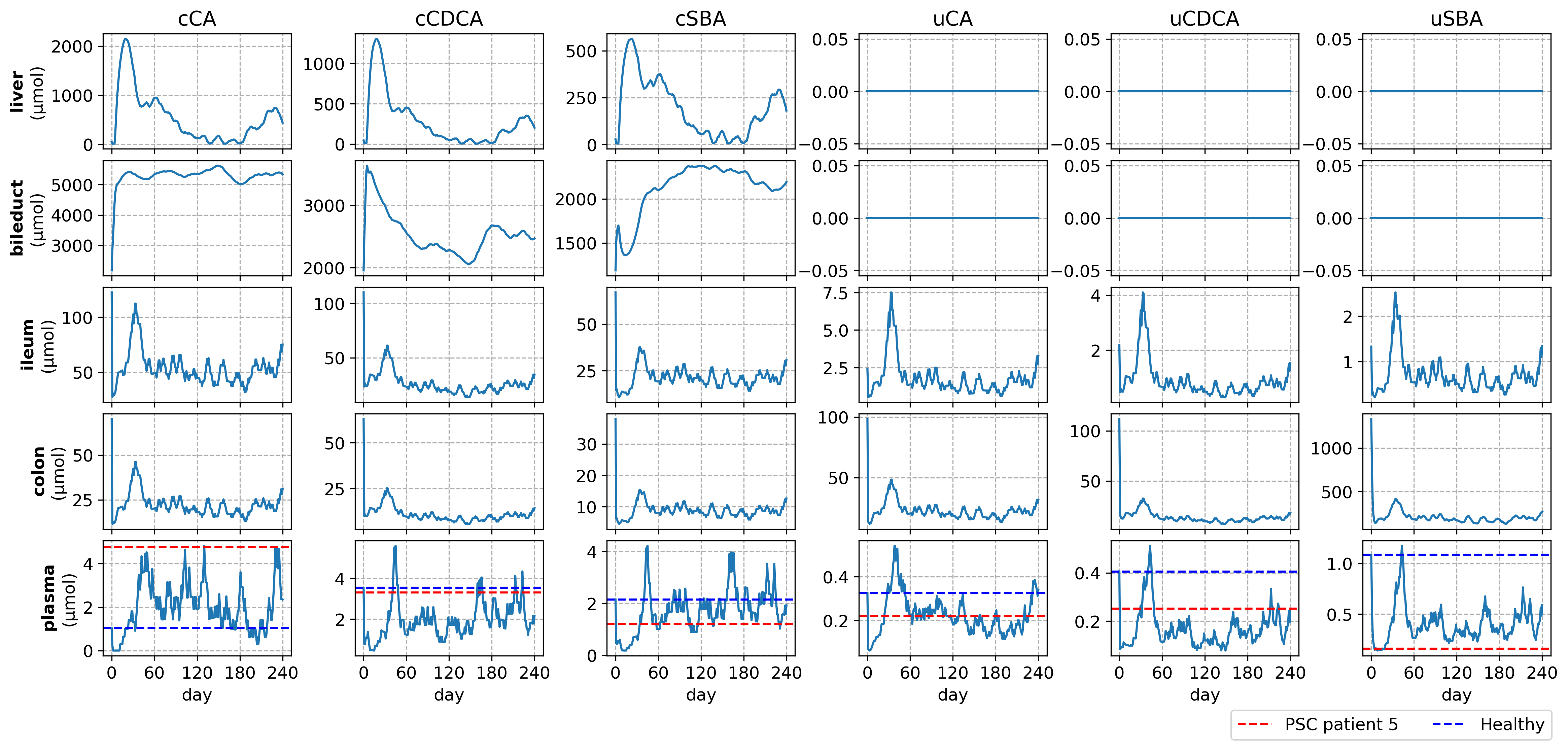}}
\end{figure*}

\begin{figure*}[htbp]
\floatconts
  {fig:psc_ode_param_w_rl_patient_5}
  {\caption{Adaptation of ODE parameters corresponding to the simulation scenario illustrated in \figureref{fig:psc_ba_w_rl_patient_5}.}}
  {\includegraphics[width=0.75\linewidth]{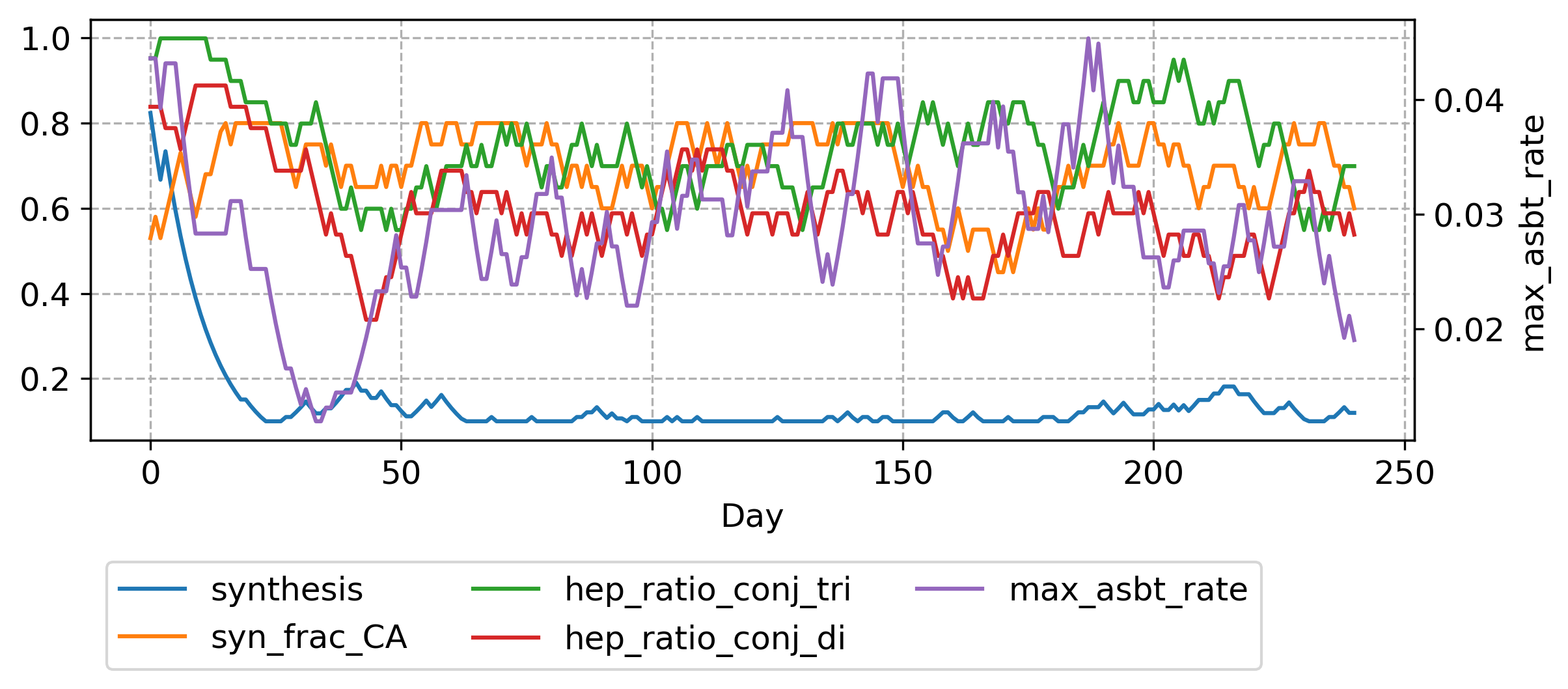}}
\end{figure*}

\end{document}